\documentclass[10pt,a4paper,reqno,oneside]{article}
\usepackage{graphics}
\usepackage{amsfonts}
\usepackage{amsmath}
\usepackage{amssymb}
\usepackage{amscd}
\usepackage{epsfig}
\usepackage{mathbbol}
\newcommand{\Om}{\Omega}
\newtheorem{theorem}{Theorem}[section]{\bf}{\it}
\newtheorem{proposition}{Proposition}[section]{\bf}{\it}
\newtheorem{lemma}{Lemma}[section]{\bf}{\it}
\newtheorem{Def}{Definition}[section]{\bf}{\it}
\newtheorem{Ass}{Assumption}[section]{\bf}{\it}
\newtheorem{Not}{Notations}[section]{\bf}{\it}
\newtheorem{Rem}{Remark}[section]{\bf}{\it}
\newcommand{\strain}{\mathcal{E}^\star}

\def\R{\mathbb{R}}

\numberwithin{equation}{section}
\usepackage[usenames]{color}
\usepackage{url}%\urlstyle{sf}
\usepackage[ps2pdf,colorlinks=true,breaklinks=true,hypertexnames=false,
            linkcolor=MyRed,citecolor=MyGreen,urlcolor=MyBlue,
            hyperindex,
            pdftitle={Thesis},
            pdfsubject={PhD Thesis},
            backref,
            pagebackref,
            filecolor=blue,
            pdfcreator = {LaTeX with hyperref package},
            pdfauthor={Nicolas Van Goethem}
            ]{hyperref}
\definecolor{MyBlue}{rgb}{0,0,0.7}
\definecolor{MyRed}{rgb}{0.7,0,0}
\definecolor{MyGreen}{rgb}{0,0.7,0}
\title{A distributional approach to the geometry of $2D$ dislocations at the mesoscale\newline-SUBMITTED PAPER-}
\author{N. Van Goethem$^1$, F. Dupret$^1$\\$^1$ CESAME, Universit\'e catholique de Louvain, \\ Louvain-la-Neuve,\ Belgium}
\date{}
\begin{document}
\maketitle
\begin{abstract}
We develop a theory to represent dislocated single crystals at the mesoscopic scale by considering concentrated effects, governed by the distribution theory combined with multiple-valued kinematic fields.
Our approach gives a new understanding of the continuum theory of defects as developed by Kr\"oner (1980) and other authors. Fundamental $2D$ identities relating the incompatibility tensor to the Frank and Burgers vectors are proved under global strain assumptions relying on the geometric measure theory, thereby giving rise to rigorous homogenisation from mesoscopic to macroscopic scale.
\end{abstract}
%
%\tableofcontents
%
\section{Introduction}
\label{intro}
Dislocations can be considered as the most undesirable and resistant class of defects for several kinds of single crystals (Maroudas and Brown, 1991; Jordan et al., 2000). Therefore, in order to improve crystal quality, the development of a relevant and accurate physical model represents a key issue with a view to reducing the dislocation density in the crystal by acting in an appropriate way on the temperature field and the solid-liquid interface shape during the growth process (Van den Bogaert and Dupret, 1997). 
\newline However the dislocation models available in the literature, such as the model of Alexander and Haasen (1968, 1986), are often based on a rather crude extension of models previously developed for polycrystals (such as usual metals and ceramics). In this case, some particular features of single crystals, such as material anisotropy or the existence of preferential glide planes, can be taken into account up to some extent, but the fundamental physics of dislocations in single crystals cannot be captured. In fact, dislocations are lines that either form loops, or end at the single crystal boundary, or join together at some locations, while each dislocation segment has a constant Burgers vector which exhibits additive properties at dislocation junctions.
These properties play a fundamental role in the modelling of line defects in single crystals and induce key conservation laws at the macro-scale (typically defined by the crystal diameter).
On the contrary, no dislocation conservation law exists at the macro-scale for polycrystals since dislocations can abruptly end at grain boundaries inside the medium without any conservation law holding across these interfaces.
\newline Aware of these principles and of the pioneer works of Volterra (1907) and Cosserat (1909),  Burgers (1939), Eshelby, Frank and Nabarro (1951, 1956), Kondo (1952), Nye (1953), and Kr\"oner (1980) among other authors (Bilby, 1960; Mura, 1987) consider a tensorial density to model dislocations in single crystals at the macro-scale, in order to take into account both the dislocation orientation and the associated Burgers vector (cf the survey contributions of Kr\"oner (1980, 1990) and Kleinert (1989)). However, in these works, the relationship between macro-scale crystal properties and the basic physics governing the nano-scale (defined by the inter-atomic distance) is not completely justified from a mathematical viewpoint. Therefore, to well define the concept of tensorial dislocation density, we here introduce an additional scale to the macro-and nano-scales, viz. the meso-scale as defined by the average distance between dislocations. The laws governing dislocation behaviour are modelled at the nano-scale, while the meso-scale (defined from the nano-scale by ensemble averaging or by averaging over a representative volume (Kr\"oner, 2001)) defines the "dislocated continuous medium", where each dislocation is viewed as a line and the interactions between dislocations can be modelled while the laws of linear elasticity govern the adjacent medium.
\newline The present paper focuses on meso-scale modelling with a view to clarifying the homogenisation process from meso- to macro-scale. Since dislocations are lines at the meso-scale, concentrated effects, as governed by the distribution theory (a key reference is here Schwartz (1957)), must be introduced in the mesoscopic model. In addition, since integration around the dislocations generates a multiple-valued displacement field with the dislocations as branching lines, multivalued functions must be considered (Almgren, 1986; Knopp, 1996; Remmert, 1996). This combination of distributional effects and multivaluedness is a key feature of the dislocation theory at the meso-scale but, unfortunately, the difficulties resulting from this mathematical association have not well been addressed so far in the literature (Thom, 1980). As an example, non-commuting differentiation operators are freely introduced without any justification by Kleinert (1989). Therefore, the principal objective of this paper is to provide a strong mathematical foundation to the meso-scale theory of dislocations, showing how the distribution and geometric measure theories can be correctly used with multiple-valued fields. In particular, the application limits of Stokes' theorem are clarified. For the sake of generality, disclinations, which represent a second but rarer kind of line defect, with in addition a multiple-valued rotation field, are here considered together with dislocations. 
\newline After homogenisation from meso- to macro-scale, no concentrated effects remain anymore present in the macroscopic model, which consists of a set of evolution PDE's governing scalar or tensorial defect density fields in the framework of elasto- or visco-plasticity (Kratochvil and Dillon, 1969). However, it should be pointed out that homogenisation from meso- to macro-scale has no meaning for multiple-valued fields such as displacement and rotation, since this operation is exclusively allowed for additive (or extensive) fields such as stress, energy density or  heat flux. This consideration becomes obvious when homogenisation is defined by an ensemble averaging procedure, since multiple-valued fields are mathematically defined as extended functions which cannot be added since their "domains" depend on the defect line locations. This issue is clarified in the present paper. Moreover, since the macroscopic displacement and rotation fields are not defined as ensemble averages of their mesoscopic counterparts, no unique privileged reference configuration can be defined at the macro-scale for single crystals with dislocations. Having in mind that displacement and rotation fields are defined with respect to the selected reference configuration (which can be, or not, defect free), the invariance laws governing the behaviour of single crystals with line defects at the macro-scale are constructed in accordance with this observation.
\newline In the literature the macroscopic dislocation density is classically defined as the curl of the plastic distortion (Head et al., 1993; Cermelli and Gurtin, 2001; Gurtin, 2002; Koslowski et al., 2002; Ariza and Ortiz, 2005), following a postulated distortion decomposition into elastic and plastic parts. However, this approach cannot be rigorously justified (contrarily to the strain decomposition) since elastic and plastic rotations cannot be set apart. In contrast, the present paper introduces the macroscopic dislocation density by homogenisation of well-defined mesoscopic fields, under precise geometric-measure model assumptions, from which the distortion decomposition is obtained together with its relationship with the dislocation density. Since dislocations and disclinations represent body torsion and curvature, respectively, these concepts also appear as macroscopic counterparts of well-defined mesoscopic defect measures.
\newline The present paper is restricted to the $2D$ theory. Extension to the $3D$ case is under investigation. A complete link between the mesoscopic and macroscopic behaviours of single crystals with line defects should be derived from these developments. In Section \ref{prelim}, the scaling analysis summarised in this introduction is detailed and the basic concepts used to represent the dislocated continuous medium are introduced. Classical invariance theorems are recalled in Section \ref{multivalued}. In Section \ref{distranal2}, the $2D$ distributional theory of the dislocated continuous medium is established in the case of a single dislocation, while Section \ref{distrset} treats the more general case of an ensemble of isolated dislocations. Finally, Section \ref{macroanal} introduces the non-Riemannian macroscopic body by homogenisation from the mesoscale, and conclusions are drawn in Section \ref{concl}.
\section{Multiscale analysis of dislocations}\label{prelim}
To address the modelling of elastic single crystals with line defects, the various scales relevant for matter description and their interrelations are briefly reviewed.
\begin{itemize}
\item At the nano-scale the characteristic length is the interatomic distance. At time $t$, the body is referred to as $\mathcal{R}^{\star\star}(t)$ and the reference body is a perfect lattice $\mathcal{R}^{\star\star}_0$.
\item At the meso-scale the characteristic length is the average distance between two neighbour dislocation lines. At time $t$, the body is referred to as ${\mathcal R}^{\star}(t)$, to be interpreted as a random sample corresponding to a given growth experiment. The reference body ${\mathcal R}^{\star}_{0}$ is a perfect crystal, i.e. a body without dislocations or disclinations.
\item At the macro-scale the characteristic length is the diameter of the crystal and the body ${\mathcal R}(t)$ has a physical meaning related to ${\mathcal R}^{\star}(t)$ and ${\mathcal R}^{\star\star}(t)$ in terms of ensemble average; the reference body ${\mathcal R}_0$ can be, or not, a perfect crystal.
\end{itemize}
\subsection{Nano-scale analysis: crystalline lattice}\label{nano}
Given a dislocation in the general sense, the atomic arrangement ${\mathcal R}^{\star\star}(t)$ differs from the reference lattice $\mathcal{R}^{\star\star}_0$, but however the atom displacements are not uniquely defined (Kleinert, 1989). More exactly, a discrete multivalued mapping $x_i:=\chi^{\star\star}_i(X_l)$ where $i=1,2$ or $3$, is defined with $X_l\in\mathcal{R}^{\star\star}_0$ and $x_i\in\mathcal{R}^{\star\star}(t)$. In general, the dislocation position cannot be determined precisely at the atomic level since several dislocation locations in the actual crystal can be associated with the same picture of the atom positions. In fact the defect should be understood as located inside a nanoscopic lattice region.
Let us insist on the fact that there is no way to uniquely define the displacement field. Indeed any atom of $\mathcal{R}^{\star\star}_0$ can in principle be selected to define the displacement of a given atom of $\mathcal{R}^{\star\star}(t)$ which is therefore a multivalued discrete mapping. This remark also makes sense at higher scales.
\subsection{Meso-scale analysis: dislocated continuous medium}\label{meso}
This scale is the one on which this paper focuses, in the framework of $2D$ linear elasticity.
Let us here describe some general and basic field properties at the meso-scale level. Details are  given in the forthcoming sections.
\begin{itemize}
\item The displacement field is a multivalued function such that for any point $X_l \in {\mathcal R}^\star_0$ one has
\begin{eqnarray}
u^\star_i(X_l)=x_i- X_i,\quad\mbox{with}\quad x_i:=\chi^\star_i(X_l),\nonumber
\end{eqnarray}
and where $\chi^\star_i(X_l)$ is a multivalued mapping from ${\mathcal R}^\star_0$ to ${\mathcal R}^\star(t)$. Displacement multivaluedness represents an important difficulty to address at the meso-scale in dislocation modelling. As opposed to multiple-valued fields, single-valued fields will also be called uniform.
\item The strain will be denoted by $\mathcal{E}^\star_{ij}$. In general, the Lagrange deformation tensor is given by
\begin{eqnarray}
\mathcal{E}^\star_{ij}:=\frac{1}{2}(\partial_j u_i^\star+\partial_i u_j^\star+\partial_j u_m^\star\partial_i u_m^\star), \quad\mbox{with}\quad \displaystyle\partial_j u_i^\star:=\frac{\partial u_i^\star}{\partial X_j},\nonumber
\end{eqnarray}
with the classical indicial notations and summation convention used.
In the sequel, linear elasticity will be assumed and hence the nonlinear terms are not taken into account at the meso-scale. This fundamental hypothesis relies on the assumption\footnote{In practise this assumption is certainly valid in single crystal growth.} that all nonlinear deformation effects take place around the dislocation in a nano-scale region whose diameter is small compared to the meso-scale characteristic distance. Therefore, using a singular perturbation asymptotic treatment, the nonlinear effects become concentrated inside the defect line at the meso-scale and hence the strain can be assumed to be the single-valued linear symmetric tensor given by
\begin{eqnarray}
\mathcal{E}_{ij}^\star:=\frac{1}{2}(\partial_j u^\star_i+\partial_i u^\star_j)\nonumber
\end{eqnarray}
outside the defect line and arbitrarily set to $0$ on the defect line (noting that concentrated deformation effects inside this line do not play any role in displacement integration at the meso-scale).
\item The infinitesimal rotation tensor is a possibly multiple-valued field given by $\omega^\star_{ij}:=\frac{1}{2}(\partial_ju^\star_i-\partial_i u^\star_j)$ with the associated rotation vector given by
\begin{eqnarray}
\omega^\star_k=-\frac{1}{2}\epsilon_{ijk}\omega^\star_{ij}=\frac{1}{2}\epsilon_{ijk}\partial_j u^\star_i\nonumber
\end{eqnarray}
and the identity $\omega^\star_{ij}=-\epsilon_{ijk}\omega^\star_k$. The Frank and Burgers vectors $\Om^\star_k$ and $B^\star_i$ associated with a defect line are commonly defined as functions of the jumps of $\omega_k^\star$ and $u_i^\star$ around this line. From Weingarten's theorems, these vectors are shown as invariants of the defect line.
\end{itemize}
The following geometric tensors are also introduced:
\begin{Def}
\begin{eqnarray}
&&\hspace{-38pt}\mbox{\scriptsize{DISCLINATION DENSITY:}}\hspace{45pt}\Theta^\star_{ij}:=\Omega^\star_j\delta_{iL},\label{disclindens1}\\
&&\hspace{-38pt}\mbox{\scriptsize{DISLOCATION DENSITY:}}\hspace{48pt}\Lambda^\star_{ij}:=B^\star_j\delta_{iL},\label{dislocdens2}\\
&&\hspace{-38pt}\mbox{\scriptsize{DISPLACEMENT JUMP DENSITY:}}\hspace{14pt}\alpha^\star_{ij}:=\Lambda^\star_{ij}+\epsilon_{jlm}\Theta^\star_{il}(x_m-x_{0m}),\label{dislocdens3}\\
&&\hspace{-38pt}\mbox{\scriptsize{CONTORTION:}}\hspace{90pt}\kappa^\star_{ij}:=\alpha^\star_{ij}-\frac{1}{2}\alpha^\star_{mm}\delta_{ij},
\label{dislocdens4}
\end{eqnarray}
where $x_{0m}$ is a reference point for rotation and displacement integration.
\end{Def}
Here, the symbol $\delta_{iL}$ is used to represent the concentrated vectorial measure density on the defect region $L$. In particular, when $L$ is a rectifiable curve, $\delta_{iL}$ is equal to $\tau_i\delta_L$ with $\tau_i$ the unit tangent vector to $L$ and $\delta_L$ the one-dimensional Hausdorff measure density concentrated on $L$.\newline
The disclination and dislocation density tensors $\Theta^\star_{ij}$ and $\Lambda^\star_{ij}$
are measure densities (cf Evans and Gariepy, 1992; Mattila, 1995) related to the so-called strain incompatibility $\eta_{ij}^\star$ to be defined later.
In general at the meso-scale a dislocation or a disclination is a defect line (i.e. a singular line for the strain) to which non-vanishing Burgers and/or Frank vectors are attached. The tensors $\Lambda^\star_{ij}$ and $\Theta^\star_{ij}$ are basic physical tools to model defect density at the meso-scale while $\eta_{ij}^\star$ plays a key role to understand their behaviour. 
The displacement jump density and mesoscopic contortion (or lattice curvature) tensors $\alpha^\star_{ij}$ and $\kappa^\star_{ij}$ are combinations of these basic density tensors, with $\alpha^\star_{ij}=\Lambda^\star_{ij}$ when the disclination density tensor vanishes.
\subsection{Macro-scale analysis: continuous medium}\label{macro}
At this level, a point $x_i$ of the actual body $\mathcal R(t)$, $x_i=\chi_i(X_l)$ where $X \in \mathcal R_0$ will be called a material point to be understood as a certain volume of matter of mesoscopic size located around the point $x_i$. In order to define macroscopic concepts such as temperature or stress, one needs to give a meaning to the temperature and stress at any point. The rigorous definition is obtained from an ergodicity argument and hence, at the macroscopic level the fields on ${\mathcal R}(t)$ are defined as ensemble averages of the fields defined on ${\mathcal R}^{\star}(t)$. By this operation these fields are smoothed, which means that concentration effects at the meso-scale level along the defect lines are erased.
To this end, a weak limiting procedure (or homogenisation) is needed in order to define the dislocation and disclination densities $\Lambda_{ij}$ and $\Theta_{ij}$
at the macro-scale level from the knowledge of the meso-scale fields $\Lambda^\star_{ij}$ and $\Theta^\star_{ij}$.
\begin{Rem}\label{trippes}
In this context, the reference body $\mathcal{R}_0$ is basically arbitrary and can, or not, be a perfect crystal. Indeed, at the macro-scale, the displacement $u_i$ must be a single-valued function, while the displacement field $u^\star_i$ is multivalued at the meso-scale. Therefore, $u^\star_i$ cannot be considered as belonging to a (linear) Banach space (single-valued functions can be added since they share the same domain, whereas a multivalued function is defined on its specific Riemann foliation and cannot be added to a multivalued function defined on another Riemann foliation). Consequently the ensemble averaging procedure is forbidden for multivalued fields such as $u_i^\star$ and hence $u_i$ is not the ensemble average of $u_i^\star$. It should also be observed that removing the field multivaluedness by performing appropriate cuts is of no use here, since by derivation these cuts introduce arbitrary distributional contributions without physical meaning. In general, it is important to make it clear that the only fields which can be obtained at the macro-scale by ensemble averaging from the meso-scale are the so-called extensive fields associated with additive physical properties (such as specific mass, stress, specific internal energy... and the dislocation and disclination densities).
\end{Rem} 
\section{Multiple-valued fields and line invariants; distributions as a modelling tool at the meso-scale}\label{multivalued}
\begin{Not}\label{notube}
In the following sections, the assumed open domain is denoted by $\Omega$ (in practise but not necessarily $\Omega$ is bounded), the defect line(s) are indicated by $L\in\Omega$, and $\Omega_L$ is the chosen symbol for $\Om\setminus L$, which is also assumed to be open without loss of generality.
Focussing on the meso-scale, symbol $\hat x$ or $\hat x_i$ denotes a generic point of the defect line(s), $x$ or $x_i$ is a generic point of $\Om_L$ and $x_0$ or $x_{0i}$ is a given fixed reference point of $\Om_L$. When $x$ and $\hat x$ are used together, $\hat x$ denotes the projection of $x$ onto the defect line(s) $L$ in a appropriate sense and $\hat{\nu}_i:=\nu_i(\hat x,x)$ is the unit vector joining $\hat x$ to $x$. The symbol $\odot_\epsilon$ is intended for a set of diameter $2\epsilon$ enclosing the region $L$. More precisely, $\odot_\epsilon$ is defined as the intersection with $\Om$ of the union of all closed spheres of radius $\epsilon$ centred on $L$: 
\begin{eqnarray}
\displaystyle\odot_\epsilon:=\Omega\cap\bigcup_{\hat x \in L}B[\hat x,\epsilon].\nonumber
\end{eqnarray} 
In case $L$ is an isolated line, $\odot_\epsilon$ is a tube of radius $\epsilon$ enclosing $L$. In the sequel, considering a surface $S$ of $\Om$ crossed by $L$ at $\hat x$ and bounded by the curve $C$, symbols $dC$, $dL$, and $dS$ will denote the $1D$ Hausdorff measures on $C$ and $L$, and the $2D$ Hausdorff measure on $S$, respectively, with $\hat\sigma_j$ and $\tau_j$ the unit tangent vectors to $C$ at $x$ and to $L$ at $\hat x$ (when they exist). In some cases (having fractal curves in mind) the symbols $dx_k$ and $dS_i:=\epsilon_{ijk}dx^{(1)}_jdx^{(2)}_k$ will stand for infinitesimal vectors oriented along $C$ and normal to $S$, respectively, with in addition $dC_l(x):=\epsilon_{lmn}dx_m\tau_n$ denoting an infinitesimal vector normal to $C$ when $\tau_n=\tau_n(\hat x)$ exists.
\end{Not}
\begin{Ass}[Mesoscopic elastic strain]\label{eta}
Henceforth we will assume that the linear strain $\mathcal{E}^\star_{mn}$ is a given symmetric $\mathcal{C}^{\infty}(\Om_L,\R^{3\times 3})$-tensor prolonged by $0$ on the line $L$, $L^1$-integrable on $\Omega$ and compatible on $\Om_L$. In other words,
the equality
\begin{eqnarray}
\epsilon_{qlm}\epsilon_{kpn}\partial_l\partial_p\mathcal{E}^\star_{mn}=0\label{eta2}
\end{eqnarray}
is assumed everywhere on $\Om_L$.
\end{Ass}
\subsection{Distributional analysis of the multiple-valued fields}\label{distranal}
In general, a multivalued function from $\Om_L$ to $\R^N$ consists of a pair of single-valued mappings with appropriate properties:\[F\rightarrow\Om_L\quad\mbox{and}\quad\ F\rightarrow \R^N,\] where $F$ is the associated Riemann foliation (Almgren, 1986; Knopp, 1996; Remmert, 1996). In the present case of meso-scale elasticity, we limit ourselves to multivalued functions obtained by recursive line integration of single-valued mappings defined on $\Om_L$. Reducing these multiple line integrals to simple line integrals, the Riemann foliation shows to be the set of equivalence path classes in $\Om_L$ from a given $x_0 \in \Om_L$ with the homotopy as equivalence relationship. Accordingly, a multivalued  function will be called of index $n$ on $\Om_L$ if its $n$-th differential is single-valued on $\Om_L$. No other kinds of multifunctions are considered in this work, whether $L$ is  a single line or belongs to a more complex set of defect lines (with possible branchings, etc.).
\begin{Not}
The notation $\partial^{(s)}_j$ is used for partial derivation of a single- or multiple-valued function whose domain is restricted to $\Om_L$. Locally around $x\in\Om_L$, for smooth functions, the meanings of $\partial^{(s)}_j$ and the classical $\partial_j$ are the same, whereas on the entire $\Om$ the partial derivation operator $\partial_j$ only applies to single-valued fields and must be understood in the distributive sense. A defect-free subset $U$ of $\Om$ is an open set such that $U \cap L=\emptyset$, in such a way that $\partial^{(s)}_j$ and $\partial_j$ coincide on $U$ for every single- or multiple-valued function of index $1$.
\end{Not}
In the following essential definition the strain is considered as a distribution on $\Omega$:
\begin{Def}\label{franktens}[Frank tensor]
The Frank tensor $\overline\partial_m\omega_k^\star$ is defined on the entire domain $\Om$ as the following distribution:
\begin{eqnarray}
\overline\partial_m\omega_k^\star:=\epsilon_{kpq}\partial_p\mathcal{E}_{qm}^\star,\label{delta_m_a}
\end{eqnarray}
in such a way that
\begin{eqnarray}
<\overline\partial_m\omega_k^\star,\varphi>:=-\int_\Om\epsilon_{kpq}\mathcal{E}_{qm}^\star\partial_p\varphi dV,\label{delta_m}
\end{eqnarray}
with $\varphi$ a smooth test-function with compact support in $\Omega$.
\end{Def}
In fact, the tensorial distribution $\overline\partial_m\omega_k^\star$ is the finite part of an integral when acting against test-functions. Indeed, since $\partial_p\mathcal{E}_{qm}^\star$ might be non-$L^1(\Om)$-integrable in view of its possibly too strong singularity near the defect line, instead of being directly calculated as an integral, $<\epsilon_{kpq}\partial_p\mathcal{E}_{qm}^\star,\varphi>$ must be calculated on $\Om$ as the limit
\begin{eqnarray}
\lim_{\epsilon\to 0}\left(\int_{\Om\setminus\odot_\epsilon}\epsilon_{kpq}\partial_p\mathcal{E}_{qm}^\star\varphi dV
+\int_{\partial\odot_\epsilon\cap\Om}\epsilon_{kpq}\mathcal{E}_{qm}^\star\varphi dS_p\right),\label{FP1}
\end{eqnarray}
where the second term inside the parenthesis is precisely added in order to
achieve convergence. One readily sees after integration by parts that expression (\ref{FP1}) is equal to Eq. (\ref{delta_m}) provided $\displaystyle \lim_{\epsilon\to 0}\Om\setminus \odot_\epsilon=\Om_L$ (which is a general hypothesis limiting the acceptable defect lines together with the assumption that $L$ is of vanishing $2D$ Hausdorff measure).
Considering the possibly multivalued (with index $1$) rotation vector $\omega^\star_k$, it should be observed from Definition \ref{franktens} that $\overline\partial_m\omega_k^\star=\partial^{(s)}_m\omega^\star_k$ on $\Om_L$. This results from the classical relationship provided by elasticity theory between infinitesimal rotation and deformation derivatives. However, $\overline\partial_m\omega_k^\star$ is defined by Eq. (\ref{delta_m_a}) as a distribution on $\Om$ and therefore concentrated effects on $L$ and its infinitesimal vicinity are added to $\partial^{(s)}_m\omega^\star_k$, justifying the use of the symbol $\overline\partial_m\omega^\star_k$ instead of $\partial_m\omega^\star_k$ without giving to $\overline\partial_m$ the meaning of a derivation operator. In particular, it may be observed that the identical vanishing of $\partial^{(s)}_m\omega^\star_k$  on $\Om_L$ does not necessarily imply that the distribution $\overline\partial_m\omega_k^\star$ vanishes as well. In fact from Eq. (\ref{FP1}), it can be shown in this particular case that
\begin{eqnarray}
<\overline\partial_m\omega_k^\star,\varphi>=\lim_{\epsilon\to 0}\int_{\partial\odot_\epsilon\cap\Om}\epsilon_{kpq}\mathcal{E}_{qm}^\star\varphi dS_p=-\int_\Om\epsilon_{kpq}\mathcal{E}_{qm}^\star\partial_p\varphi dV,
\end{eqnarray} 
which is generally non-vanishing. Finally, as soon as the definition of the tensor distribution $\overline\partial_m\omega_k^\star$ is given, so are the distributional derivatives of $\overline\partial_m\omega_k^\star$:
\begin{eqnarray}
<\partial_l\overline\partial_m\omega_k^\star,\varphi>=-<\overline\partial_m\omega_k^\star,\partial_l\varphi>
=\int_\Om\epsilon_{kpn}\mathcal{E}_{mn}^{\star}\partial_p\partial_l\varphi dV.\label{FP2}
\end{eqnarray}
\subsection{Rotation and displacement vectors}\label{rotdispl}
The rotation vector is defined from the knowledge of the linear strain together with the rotation at a given reference point $x_0$. From this construction follows an invariance property of $\omega^\star_k$ as a multifunction (recalling that  multivaluedness takes its origin from the existence of a defect line which renders the strain incompatible on the entire $\Om$).\newline
Starting from the distributive Definition \ref{franktens} of $\overline\partial_m\omega_k^\star$, the differential form $\overline\partial_m\omega^\star_kd\xi_m$ is integrated along a regular parametric curve $\Gamma \subset \Om_L$ with endpoints $x_0, x\in \Om_L$. For selected $x_0$ and $\omega^\star_{0k}$, the multivalued rotation vector is defined as
\begin{eqnarray}
\omega^\star_k=\omega^\star_k(\#\Gamma,\omega^\star_0)=\omega^\star_{0k}+\int_\Gamma\overline\partial_m\omega^{\star}_kd\xi_m,\nonumber
\end{eqnarray}
where $\#\Gamma$ is the equivalence class of all regular curves homotopic to $\Gamma$ in $\Om_L$. Indeed, from strain compatibility in $\Om_L$, i.e. from relation (\ref{eta2}), it is clear that $\omega^\star_k$ is a function of $\#\Gamma$ only. Consider now a regular parametric loop $C$ (in case $C$ is a planar loop, it is called Jordan curve) and the equivalence class $\#C$ of all regular loops homotopic to $C$ in $\Om_L$. Here, the extremity points play no role anymore and two loops are equivalent if and only if they can be continuously transformed into each other in $\Om_L$. The jump of the rotation vector $\omega^\star_k$ along $\#C$ depends on $\#C$ only and is defined as\footnote{We note that the curve $C$ could be non rectifiable, i.e. of infinite length. In fact, integrals on fractal curves and the related Stokes' and Gauss-Green's theorems are analysed by Harrison and Norton (1992), where it is shown, by the $\mathcal{C}^\infty$-smoothness of the differential form $\overline\partial_m \omega^\star_k dx_m$ on $\Om_L$ that Eq. (\ref{GG}) still holds even when the Hausdorff dimension of $C$ is higher than $1$.}
\begin{eqnarray}
[\omega^\star_k]=[\omega^\star_k](\#C)=\int_C \overline\partial_m \omega^\star_k d\xi_m.\label{GG}
\end{eqnarray}
The following developments address the displacement field  multivaluedness as a mere consequence of strain incompatibility. The procedure defining the displacement vector from the rotation vector by means of line integrals is classical in linear elasticity. The following tensor plays in the construction of the displacement field a role analogous to $\overline\partial_m\omega^\star_k$ in the construction of the rotation field:
\begin{Def}\label{burgtens}[Burgers tensor]
For a selected reference point $x_0\in\Om_L$, the Burgers tensor is defined on the entire domain $\Om$ as the distribution
\begin{eqnarray}
\displaystyle\overline\partial_l b^\star_k(x;x_0):=\mathcal{E}^\star_{kl}(x)+\epsilon_{kpq}(x_p-x_{0p})\overline\partial_l\omega^\star_q(x).\label{delta_lb_i}
\end{eqnarray}
\end{Def}
The Burgers tensor can be integrated in the same way as the Frank tensor along any parametric curve $\Gamma$, providing for selected $x_0$, $\omega^\star_{0k}$ and $u^\star_{0k}$ the multivalued displacement vector $u^\star_k$ of index $2$:
\begin{eqnarray}
u_k^\star=u_k^\star(\#\Gamma,\omega^\star_0,u^\star_0)=u^\star_{0k}+\epsilon_{klm}\omega_l^\star(x;\Gamma)(x_m-x_{0m})+\int_\Gamma\overline\partial_l b^\star_k(\xi)d\xi_l,\nonumber
\end{eqnarray}
which is a function of $\#\Gamma$ only. It may be observed that $\overline\partial_l b^\star_k$ and the vector $b^\star_k=u^\star_k-\epsilon_{klm}\omega_l^\star(x_m-x_{0m})$ are related in the same way as $\overline\partial_m\omega^\star_k$ and $\omega^\star_k$, including the fact that $\overline\partial_l b^\star_k=\partial_l^{(s)} b^\star_k$ on $\Om_L$. The jumps of $b^\star_k$ along $\#C$ and of $u^\star_k$ at $x$ along $\#C$ (which depends on $\#C$ only) are defined as
\begin{eqnarray}
[b^\star_k](\#C;x_0)=[u_k^\star](x;\#C;x_0)-\epsilon_{klm}[\omega^\star_l](\#C)(x_m-x_{0m})=\int_C \overline\partial_lb^\star_k d\xi_k.
\end{eqnarray}
Let us now, for the sake of simplicity, focus on the case of a defect line $L$ which $(i)$ can itself be represented as a single $\mathcal{C}^0$, closed or not, parametric line without multiple points except possibly its extremities and $(ii)$  is isolated in the sense that each of its points $\hat x$ is located inside a smooth surface $S(\hat x)$ bounded by a loop $C(\hat x)$ and such that $S(\hat x)\setminus\{\hat x\}\subset \Om_L$. Such a defect line $L$ will be called an isolated dislocation or disclination. The jump $[\omega^\star_k]$ of the rotation vector $\omega^\star_k$ around $L$ is defined as the jump of $\omega^\star_k$ along $\#C(\hat x)$ and hence is the same for any $\hat x$ and suitable $C(\hat x)$. Similarly, the jump $[b^\star_k]$ of the vector $b^\star_k$ around $L$ is defined as the jump of $b^\star_k$ along $\#C(\hat x)$ and is also the same for any $\hat x$ and suitable $C(\hat x)$, given $x_0$. In fact, the following result is well-known (Kleinert, 1989):
\begin{theorem}\label{Wein1}[Weingarten]
The rotation vector $\omega_k^{\star}$ is a multifunction of index $1$ on $\Om_L$ whose jump  $\Om^{\star}_k:=[\omega_k^{\star}]$ around $L$ is an invariant of the defect-line $L$. Moreover, for a given $x_0$, the vector $b_k^\star$ is a multifunction of index $1$ on $\Om_L$ whose jump  $B^\star_k:=[b_k^\star]$ around $L$ is an invariant of the defect-line.
\end{theorem}
\begin{proposition}\label{displ}[Multiple-valued displacement field]
>From a symmetric smooth linear strain tensor $\mathcal{E}^{\star}_{ij}$ on $\Om_L$ and a point $x_0$ where the displacement and rotation are given, a multivalued displacement field $u^\star_i$ of index $2$ can be constructed on $\Om_L$ such that the symmetric part of the deformation gradient $\partial^{(s)}_ju^\star_i$ is the single-valued strain tensor $\mathcal{E}^\star_{ij}$ on $\Om_L$ while its skew-symmetric part is the multivalued tensor $\omega^\star_{ij}:=-\epsilon_{ijk}\omega^\star_k$.
\end{proposition}
>From this result, the Frank and Burgers vectors can be defined as invariants of the single isolated line $L$.
\begin{Def}\label{Burgers}[Frank and Burgers vectors]
The Frank vector of the line $L$ is the invariant
\begin{eqnarray}
\Om^\star_k:=[\omega^\star_k],\label{frank}
\end{eqnarray}
while for a given reference point $x_0$ its Burgers vector is the invariant
\begin{eqnarray}
B^\star_k:=[b^\star_k]=[u_k^\star](x)-\epsilon_{klm}\Om^\star_l(x_m-x_{0m}).\label{burgers}
\end{eqnarray}
A defect line with non-vanishing Frank vector is called a disclination while a defect line with non-vanishing Burgers vector is called a dislocation.
\end{Def}
Clearly a disclination should always be considered as a dislocation by appropriate choice of $x_0$ while the reverse statement is false since $\Om^\star_k$ might vanish. This is why in the present paper, the word "dislocation" means in the general sense a dislocation and/or a disclination. A pure dislocation is a dislocation with vanishing Frank vector. 
\begin{Rem}
Two distinct reference points $x_0$ and $x'_0$ define two distinct Burgers vectors, related by
\begin{eqnarray}
B^\star_k-B'^\star_k=\epsilon_{klm}(x_{0m}-x'_{0m})\Om^\star_l,\nonumber
\end{eqnarray}
in such a way that $B^\star_k\Om_k$ is an invariant independent of the arbitrary choice of $x_0$. Therefore, for a non-zero Frank vector, the vanishing of the Burgers vector depends on the arbitrary choice of $x_0$.
\end{Rem}
The following result can be readily shown and is fundamental in the framework of our investigations since it implies conservation laws at the meso- and macro-scales.
\begin{theorem}
Single disclination and dislocation lines are always closed or end at the boundary of $\Om$. Moreover,
\begin{eqnarray}
\partial_i\Theta_{ij}^\star&=&0\nonumber\\
\partial_i\alpha_{ij}^\star&=&-\epsilon_{jmn}\Theta_{mn}^\star.\nonumber
\end{eqnarray}
\end{theorem}
\begin{Def}[Mesoscopic strain incompatibility]\label{incompatibility1}
According to Eq. (\ref{eta2}) combined with Eq. (\ref{delta_m}), the incompatibility tensor is defined by
\begin{eqnarray}
\eta^{\star}_{lk}:=\epsilon_{lmn}\partial_m\overline\partial_n\omega^\star_k.\nonumber
\end{eqnarray}
The strain field is called compatible on the set $U$ if the associated incompatibility tensor vanishes on $U$.
\end{Def}
\section{Distributional analysis of incompatibility for a single rectilinear dislocation}\label{distranal2}
\subsection{The $2D$ model for rectilinear dislocations}\label{2Dsection}
$2D$ elasticity means that the strain $\mathcal{E}^\star_{ij}$ is independent of the "vertical" coordinate $z$. However this assumption introduces no restriction on the dependence of the multiple-valued displacement and rotation fields upon $z$. 
\begin{Not}\label{planarnot}
Henceforth the single defect line will be assumed to be located along the $z$-axis.
The two planar coordinates will be denoted by $x$ and $y$ or $x_\alpha$. The projection of $x=(x_\alpha,z)$ on $L$ is $\hat x=(0,0,z)$. By convention, Latin indices $i,j,k,l,\cdots$ take their values from $1$ to $3$ and are basically used for $3D$ elasticity, while Greek indices $\alpha,\beta,\gamma,\delta,\cdots$ take the values $1$ or $2$ and are used for $2D$ elasticity. Symbols $(e_x,e_y,e_z)$ or $(e_\alpha,e_z)$ denote the Cartesian base vectors, while $(e_r,e_\theta,e_z)$ denote the local cylindrical base vectors. For a planar curve $C$, the notation $dC_\alpha(x)=\epsilon_{\alpha\beta}dx_\beta$ will be used for the curve normal.
\end{Not}
Let us observe that many fields are singular at the origin and that $\Om_L$ is in fact the domain where the laws of linear elasticity apply. Moreover, the strain can  be decomposed into three tensors:
\begin{eqnarray}
\mathcal{E}^\star_{ij}=\underbrace{\delta_{\alpha i}\delta_{\beta j}\mathcal{E}^\star_{\alpha\beta}}_{planar\ strain}+\underbrace{\left(\delta_{i z}\delta_{j \gamma}\mathcal{E}^\star_{\gamma z}+\delta_{j z}\delta_{i \gamma}\mathcal{E}^\star_{\gamma z}\right)}_{3D\ shear}\underbrace{+\delta_{i z}\delta_{j z}\mathcal{E}^\star_{zz}.}_{pure\ vertical\ compression}\nonumber
\end{eqnarray}
The following propositions can be readily proved from Assumption \ref{eta}: 
\begin{proposition}\label{complane}[$2D$ compatibility]
In $\Om_L$, from $2D$ strain compatibility, there are real numbers $K, a_\alpha$ and $b$ such that
\begin{eqnarray}
\left\{\begin{array}{lll}\epsilon_{\alpha\gamma}\epsilon_{\beta\delta}\partial_\alpha\partial_\beta\mathcal{E}_{\gamma\delta}^\star=0, \\
\epsilon_{\alpha\beta}\partial_\alpha\mathcal{E}_{\beta z}^\star=K, \\ \mathcal{E}_{zz}^\star=a_\alpha x_\alpha+b.\end{array}\right.\label{eqcomplane}
\end{eqnarray}
\end{proposition}
\begin{lemma}\label{fleurs}
Let $C(\hat x)$ denote a family of $2D$ closed rectifiable curves. Then, in $2D$ elasticity, the Frank tensor and the strain verify the relation
\begin{equation}
\lim_{C(\hat x)\to\hat x}\int_{C(\hat x)}x_\alpha\overline\partial_\beta\omega_\kappa^\star dx_\beta +\epsilon_{\kappa\beta}\mathcal{E}_{\beta z}^\star dx_\alpha=0,\nonumber
\end{equation}
provided the length of $C$ is uniformly bounded and as long as the convergence $C(\hat x)\to\hat x$ is understood in the Hausdorff sense, i.e. in such a way that
\begin{eqnarray}
\max\{\Vert x-\hat x\Vert, x\in C(\hat x)\}\to 0.\nonumber
\end{eqnarray}
\end{lemma}
{\bf Proof.} 
The second compatibility condition of Eq. (\ref{eqcomplane}) is equivalent to
\[\partial_\gamma\mathcal{E}_{\beta z}^\star-\partial_\beta\mathcal{E}_{\gamma z}^\star=K\epsilon_{\gamma\beta},\]
from which, so far as $2D$ elasticity is concerned,
\begin{eqnarray}
\overline\partial_\beta\omega^\star_\kappa:=\epsilon_{\kappa\gamma}\partial_\gamma\mathcal{E}_{\beta z}^\star=\epsilon_{\kappa\gamma}\partial_\beta\mathcal{E}_{\gamma z}^\star-K\delta_{\kappa\beta},\nonumber
\end{eqnarray}
and
\begin{eqnarray}
\left(x_\alpha\overline\partial_\beta\omega_\kappa^\star+\delta_{\alpha\beta}\epsilon_{\kappa\gamma}\mathcal{E}_{\gamma z}^\star\right)=\partial_\beta\Bigl(x_\alpha\epsilon_{\kappa\gamma}
\mathcal{E}_{\gamma z}^\star\Bigr)-x_\alpha K\delta_{\kappa\beta}.\nonumber
\end{eqnarray}
Since, under the limit assumptions of this lemma,
\begin{eqnarray}
\lim_{C(\hat x)\to \hat x}\int_{C(\hat x)}x_\alpha dx_\kappa=0,\nonumber
\end{eqnarray}
and since the strain is a single-valued tensor, the proof is achieved.{\hfill $\square$}

\begin{lemma}\label{racine}
In $2D$ elasticity the planar Frank vector $\Om^\star_\alpha$ vanishes.
\end{lemma}
{\bf Proof.} 
Since
\begin{eqnarray}
\overline\partial_\beta b^\star_\tau=\mathcal{E}^\star_{\beta\tau}+\epsilon_{\tau\gamma}(x_\gamma-x_{0\gamma})\delta_\beta\omega^\star_z
-\epsilon_{\tau\gamma}(z-z_0)\delta_\beta\omega^\star_\gamma,\nonumber
\end{eqnarray}
the planar Burgers vector simply writes as
\begin{eqnarray}
B^\star_\tau=\int_{C}\left(\mathcal{E}^\star_{\beta\tau}+\epsilon_{\tau\gamma}(x_\gamma-x_{0\gamma})\delta_\beta\omega^\star_z\right)dx_\beta
-\epsilon_{\tau\gamma}(z-z_0)\Omega^\star_\gamma,\nonumber
\end{eqnarray}
where $C$ is any planar loop.
By Weingarten's theorems the Burgers vector is a constant while the integrand is independent of $z$, from which the result obviously follows.{\hfill $\square$}
\newline\newline
In general, the present theory does not make any use of the linear elasticity constitutive laws and of the momentum and energy conservation laws, since in the framework of Continuum Mechanics arbitrary body forces and heat supply could be applied. However the remaining of this section will be devoted to present the three classical examples of $2D$ line-defects for which the medium is assumed to be body force free and isothermal (detailed computations are given in Appendix \ref{2Dsection}).
\begin{itemize}
\item \textit{Pure screw dislocation.} From Eq. (\ref{displvert}), (\ref{rotvert}), and Proposition \ref{solutions}, the displacement and rotation vectors write as
\begin{eqnarray}
u^\star_ie_i=\frac{B^\star_z\theta}{2\pi}e_z\quad\mbox{and}\quad \omega^\star_ie_i=\frac{1}{2}\nabla\times u^\star_ie_i=\frac{B^\star_z}{4\pi r}e_r,\label{disp_screw}
\end{eqnarray}
in such a way that the jump $[\omega^\star_i]$ vanishes identically, while from Eq. (\ref{strainvert}) the Cartesian strain writes as
\begin{eqnarray}
[\mathcal{E}^\star_{ij}]=\frac{-B^\star_z}{4\pi r^2}\left[ \begin{array}{ccc} 0& 0 & y\\ 0 & 0 & -x \\ y & -x & 0
\end{array} \right].\label{strain_screw}
\end{eqnarray}
Moreover, in $\Om_L$, appealing to Eq. (\ref{strain_screw}), the Frank tensor writes as 
\begin{eqnarray}\label{frank_screw}
[\overline\partial_m\omega^\star_k]=\frac{-B^\star_z}{4\pi r^{2}}\left[\begin{array}{ccc} \cos2\theta & \sin2\theta & 0 \\ \sin2\theta &-\cos2\theta   & 0 \\ 0 & 0 & 0
\end{array}\right].
\end{eqnarray}
\item \textit{Pure edge dislocation.} From Eq. (\ref{displ-rot_1}), (\ref{displ-rot_2}), and Proposition \ref{solutions}, the displacement is the vector \[u^\star_i e_i=\frac{-B^\star_y(\log\frac{r}{R}+1)}{2\pi}e_x+\frac{B^\star_y\theta}{2\pi}e_y,\]
while the rotation $\omega^\star_i$ vanishes together with its jump. The Cartesian strain writes from Eq. (\ref{strain}) as
\begin{eqnarray}
[\mathcal{E}^\star_{ij}]=\frac{-B^\star_y}{2\pi r^{2}}\left [ \begin{array}{ccc} x&y& 0 \\ y&-x  & 0  \\ 0 & 0 &0
\end{array} \right ],\label{strain_edge}
\end{eqnarray}
noting that the tensor $\overline\partial_m\omega_k^\star=\nabla\omega$ vanishes identically in $\Om_L$.
\item \textit{Wedge disclination.} From Eqs. (\ref{displ-rot_2}), (\ref{displ-rot_1}), and Proposition \ref{solutions}, the rotation vector is \[\omega^\star_ie_i=\frac{\Om^\star_z\theta}{2\pi}e_z,\]
with the multiple-valued planar displacement field given by
\begin{eqnarray}
&u^\star_x-iu^\star_y=\frac{\Om^\star_z}{4\pi}(1-\nu^*)x\ln(\frac{r}{R})-\frac{\Om^\star_z}{8\pi}(1+\nu^*)x-\frac{\Om_z^\star}{2\pi}y\theta&\nonumber\\
&-i\left[\frac{\Om^\star_z}{4\pi}(1-\nu^*)y\ln(\frac{r}{R})-\frac{\Om^\star_z}{8\pi}(1+\nu^*)x+\frac{\Om^\star_z}{2\pi}x\theta\right]\label{u_disclin}
\end{eqnarray}
and a vanishing Burgers vector: \[B^\star_x-iB^\star_y=[u^\star_x]-i[u^\star_y]+\Om^\star_z(y+ix)=0.\]
The Cartesian strain writes from Eq. (\ref{strain}) as
\begin{eqnarray}
[\mathcal{E}^\star_{ij}]&=&\frac{\Om_{z}(1-\nu^{\star})}{4\pi}\left [ \begin{array}{ccc} (\log\frac{r}{R}+1)& 0& 0 \\0&(\log\frac{r}{R}+1) & 0  \\ 0 & 0 &0
\end{array} \right ]\nonumber
\\&-&\frac{\Om^\star_z(1+\nu^*)}{8\pi}\left [ \begin{array}{ccc} \cos2\theta& \sin2\theta & 0 \\\sin2\theta &-\cos2\theta & 0  \\ 0 & 0 &0
\end{array} \right]\label{strain_disclin},
\end{eqnarray}
and hence
\begin{eqnarray}
[\overline\partial_m\omega^\star_k]-\frac{\Om^\star_{z}}{2\pi r}\left[\begin{array}{ccc} 0& 0& \sin\theta \\0&0&-\cos\theta \\ 0 & 0 &0
\end{array}\right].\label{frank_disclin}
\end{eqnarray}
\end{itemize}
\begin{Rem}
It should be noted that the energy density (or compliance) $\mathcal{E}^\star=\frac{1}{2}\sigma_{ij}^\star\mathcal{E}^\star_{ij}$ is not $L^1$-integrable for both kinds of dislocations, while it is finite for the wedge disclination. Therefore, a Hadamard finite part distribution (Schwartz, 1957; Estrada and Kanwal, 1989) is needed to represent the compliance at the meso-scale (another approach makes use of strain mollification by a so-called core tensor (Koslowski et al., 2002)). This issue, whose solution requires to develop matched asymptotic expansions around the singular line in accordance with the infinitesimal displacement hypothesis, will not be addressed further in the present paper which only focuses on the geometry of dislocations.
\end{Rem}
\subsection{Mesoscopic incompatibility for a single defect line}\label{mesosingle}
For $2D$ problems the incompatibility vector contains all the information provided by the general incompatibility tensor. The latter expresses on the one hand the non-commutative action of the defect line over the second derivatives of the rotation vector and on the other hand is related to concentrated effects of the Frank and Burgers vectors along the defect line.
\begin{Def}\label{incompatibility}
In the $2D$ case, the mesoscopic incompatibility vector is defined by
\begin{eqnarray}
\eta^{\star}_k:=\epsilon_{\alpha\beta}\partial_\alpha\overline\partial_\beta\omega^{\star}_k.\label{eta-om}
\end{eqnarray}
A strain field is compatible if the associated incompatibility vector vanishes.
\end{Def}
As shown in the following sections, concentration effects will be represented by means of first- and second-order distributions.
\begin{Not} Recalling Notations \ref{planarnot}, $\Om_z$ and $\Om_z^0$ stand for the sets $\Om_z:=\{x\in\Om\quad\mbox{such that}\quad x=(x_\alpha,z)\}$ and $\Om_z^0:=\Om_z\setminus L$, while the radius $r=\Vert x-\hat x\Vert$ is the distance from a point $x$ inside $\Om$ to $L$. Then, the $1D$ Hausdorff measure concentrated on $L$ is denoted by $\delta_L$ (cf Ambrosio et al. (2000), Evans and Gariepy (1992) and Mattila (1995) for general definitions and properties on the geometric measure theory).
\end{Not}
In what follows the hypothesis consists in assuming that the strain radial dependence in the vicinity of $L$ is less singular than a critical threshold. This is verified for instance by the wedge disclination whose strain radial behaviour is $O(\ln r)$ and by the screw and edge dislocations whose strains are $O(r^{-1})$.\footnote{A function $f(\epsilon)$ is said to be $O\left(g(\epsilon)\right)(\epsilon\to 0^+)$ if there exists $K, \epsilon_0>0$ s.t. $\displaystyle 0<\epsilon<\epsilon_0\Rightarrow |f(\epsilon)|\leq K|g(\epsilon)|$. A function $f(\epsilon)$ is said to be $o\left(g(\epsilon)\right)(\epsilon\to 0^+)$ if $\displaystyle \lim_{\epsilon\to 0^+}\frac{f(\epsilon)}{g(\epsilon)}=0$.}
For a straight defect-line $L$, according to these examples, the hypotheses on the strain and Frank tensors read as follows:
\begin{Ass}\label{asstrain} [$2D$ strain for line defects]
The strain tensor $\mathcal{E}_{ij}^\star$ is independent of the vertical coordinate $z$, is compatible on $\Om_L$ in the sense that conditions (\ref{complane}) hold, is smooth on $\Om_L$ and $L^1$-integrable on $\Omega$. 
\end{Ass}
\begin{Ass}\label{asstrainloc} [Local behaviour]
The strain tensor $\mathcal{E}_{ij}^\star$ is assumed to be $o(r^{-2})$ $(\epsilon\to 0^+)$ while the Frank tensor is assumed to be $o(r^{-3})(\epsilon\to 0^+)$.
\end{Ass}
\begin{theorem}\label{mainresultloc}[Main result for a single line]
Under Assumption \ref{asstrain} and \ref{asstrainloc}, for a dislocation located along the $z$-axis, incompatibility as defined by Eq. (\ref{eta-om}) is the vectorial first order distribution
\begin{eqnarray}
\eta_k^{\star}=\delta_{kz}\eta_z^{\star}+\delta_{k\kappa}\eta_\kappa^{\star},\label{eta_k}\nonumber
\end{eqnarray}
where its vertical component is
\begin{eqnarray}
\eta^\star_z=\Om^\star_z\delta_L+\epsilon_{\alpha\gamma}\left(B^\star_\gamma-\epsilon_{\beta\gamma}x_{0\beta}\Om^\star_z\right)\partial_\alpha\delta_L,\label{final1}
\end{eqnarray}
while its planar components are
\begin{eqnarray}
\eta^\star_\kappa=\frac{1}{2}\epsilon_{\kappa\alpha}B^\star_z\partial_\alpha\delta_L\label{final2}.
\end{eqnarray}
\end{theorem}
{\bf Proof.} 
For some small enough $\epsilon>0$ and using Notations \ref{notube} a tube $\odot_\epsilon$ can be constructed around $L$ and inside $\Om$. Assuming that the smooth 3D test-function $\varphi$ has its compact support containing part of $L$, $\Om_{\epsilon,z}$ denotes the slice of $\Om\setminus\odot_\epsilon$ obtained for a given $\hat x\in L$, i.e.
\begin{equation}
\Om_{\epsilon,z}:=\{x\in\Om_z\quad\mbox{such that}\quad ||x_\alpha||>\epsilon\},\nonumber
\end{equation}
while the boundary circle of $\Om_{\epsilon,z}$ is designated by $C_{\epsilon,z}$.
\newline 
$\blacktriangle$ Let us firstly treat the left-hand side of Eq. (\ref{eta_k}). Indeed, from Definition \ref{incompatibility} with Eq. (\ref{delta_m_a}), Definition \ref{franktens}, and Eqs. (\ref{delta_m}) and (\ref{FP1}), it follows that
\begin{equation}\label{marron}
<\eta^\star_k,\varphi>=\int_Ldz\lim_{\epsilon\to 0^+}\Pi_k(z,\varphi,\epsilon),\nonumber
\end{equation}
where
\begin{equation}\label{lefthand}
\Pi_k(z,\varphi,\epsilon):=-\int_{\Om_{\epsilon,z}}\epsilon_{\alpha\beta}\overline\partial_\beta\omega^\star_k\partial_\alpha
\varphi dS-\int_{C_{\epsilon,z}}\epsilon_{\alpha\beta}\epsilon_{k\gamma n}\mathcal{E}_{\beta n}^\star\partial_\alpha\varphi dC_\gamma.\nonumber
\end{equation}
The boundedness of $|\partial_\tau\partial_\delta\varphi|$ on $\Om_L$ provides the following Taylor expansions of $\varphi$ and of $\partial_\alpha\varphi$ around $\hat x$:
\begin{eqnarray}
\varphi(x)&=&\varphi(\hat x)+r\hat{\nu}_\alpha\partial_\alpha\varphi(\hat x)+\frac{r^2}{2}\hat{\nu}_\tau\hat{\nu}_\delta\partial_\tau
\partial_\delta\varphi\left(\hat x+\gamma_1(x-\hat x)\right),\label{taylor1}\\
\partial_\alpha\varphi(x)&=&\partial_\alpha\varphi(\hat x)+r\hat{\nu}_\tau\partial_\tau
\partial_\alpha\varphi\left(\hat x+\gamma_2(x-\hat x)\right),\label{taylor2}
\end{eqnarray}
with $0<\gamma_1(x-\hat x),\gamma_2(x-\hat x)\leq 1$.\newline
$\blacktriangle$ Consider the first term of Eq. (\ref{lefthand}), noted $\hat\Pi_k$. By virtue of strain compatibility on $\Om_L$ and Gauss-Green's theorem, this term writes as
\begin{equation}
\hat\Pi_k(z,\varphi,\epsilon):=-\int_{\Om_{\epsilon,z}}\partial_\gamma\left(\epsilon_{\gamma\beta}\overline\partial_\beta\omega^{\star}_k\varphi\right)dS=\int_{C_\epsilon}\epsilon_{\gamma\beta}\overline\partial_\beta\omega^\star_k\varphi dC_\gamma.\nonumber
\end{equation}
Since by Notations \ref{planarnot} $r\hat\nu_\alpha:=x_\alpha-\hat x_\alpha=x_\alpha$, then Eq. (\ref{taylor1}) and Assumption \ref{asstrainloc} show that, for $\epsilon\to 0^+$,
\begin{eqnarray} 
\hat\Pi_k=\int_{C_{\epsilon,z}}\epsilon_{\gamma\beta}\overline\partial_\beta\omega^\star_k\Bigl(\varphi(\hat x)+x_\alpha\partial_\alpha\varphi(\hat x)\Bigr)dC_\gamma+o(1).\nonumber
\end{eqnarray}
$\blacktriangle$ Consider the second term of Eq. (\ref{lefthand}), noted $\Pi_k^\star$. On account of Assumption \ref{asstrainloc} and by expansion (\ref{taylor2}), this term may be rewritten as
\begin{eqnarray}
\Pi_k^\star(z,\varphi,\epsilon)&:=&-\int_{C_{\epsilon,z}}\epsilon_{\alpha\beta}\epsilon_{k\gamma n}\mathcal{E}_{\beta n}^\star\partial_\alpha\varphi dC_\gamma\nonumber\\
&=&-\partial_\alpha\varphi(\hat x)\int_{C_{\epsilon,z}}\epsilon_{\alpha\beta}\epsilon_{k\gamma n}\mathcal{E}_{\beta n}^\star dC_\gamma+o(1).\nonumber
\end{eqnarray}
$\blacktriangle$ From Weingarten's theorem, recalling that $dC_\gamma=\epsilon_{\gamma\tau}dx_\tau$, the expression $\Pi_k=\hat{\Pi}_k+\Pi_k^\star$ writes as
\begin{eqnarray}
\Pi_k&=&\partial_\alpha\varphi(\hat x)\int_{C_{\epsilon,z}}\left(x_\alpha
\overline\partial_\tau\omega^\star_k-\epsilon_{\alpha\beta}\epsilon_{k\gamma n}\epsilon_{\gamma\tau}\mathcal{E}_{\beta n}^\star\right)dx_\tau\nonumber\\&+&\Om^\star_k\varphi(\hat x)+o(1).\label{pik}
\end{eqnarray}
$\blacktriangle$ Consider the first term of Eq. (\ref{pik}), noted $\Pi'_k$, and take $\delta=\gamma$ in the identity
\begin{eqnarray}
\epsilon_{k\delta n}\epsilon_{\gamma\tau}=\delta_{kz}\left(\delta_{\gamma\delta}\delta_{n\tau}-\delta_{n\gamma}\delta_{\tau\delta}\right)-\delta_{nz}\left(\delta_{\gamma\delta}
\delta_{k\tau}-\delta_{k\gamma}\delta_{\tau\delta}\right)\label{identity}
\end{eqnarray}
in such a way that
\begin{eqnarray}
\Pi'_k\partial_\alpha\varphi(\hat x)\int_{C_{\epsilon,z}}\left(x_\alpha
\overline\partial_\tau\omega^\star_k-\delta_{kz}\epsilon_{\alpha\beta}
\mathcal{E}_{\beta\tau}^\star+\delta_{k\tau}\epsilon_{\alpha\beta}
\mathcal{E}_{\beta z}^\star\right)dx_\tau.\label{ombre}
\end{eqnarray}
$\blacktriangle$ The cases $k=z$ and $k=\kappa$ are treated separately.
\begin{itemize}
\item When $k=z$, Definition \ref{burgtens} shows that
\begin{eqnarray}
\displaystyle\overline\partial_\beta b^\star_\tau:=\mathcal{E}^\star_{\beta\tau}+\epsilon_{\tau\gamma}(x_\gamma-x_{0\gamma})\overline\partial_\beta\omega^\star_z-\epsilon_{\tau\gamma}(z-z_0)\overline\partial_\beta\omega^\star_\gamma\nonumber
\end{eqnarray}
which, after multiplication by $\epsilon_{\tau\alpha}$ and using Eq. (\ref{identity}) with $\tau,\alpha$ and $z$ substituted for $k,\delta$ and $n$, is inserted into Eq. (\ref{ombre}), thence yielding:
\begin{eqnarray}
\Pi'_z=\partial_\alpha\varphi(\hat x)\int_{C_{\epsilon,z}}
\left(\epsilon_{\tau\alpha}\overline\partial_\beta b_\tau^\star+x_{0\alpha}\overline\partial_\beta\omega_z^\star
+(z-z_0)\overline\partial_\beta\omega_\alpha^\star\right)dx_\beta,\label{mousse}
\end{eqnarray}
and consequently, from the definitions of the Frank and Burgers vectors,
\begin{eqnarray}\label{feuilles}
\lim_{\epsilon\to 0^+}\Pi'_z=\ \ll\left\{\epsilon_{\alpha\tau}B_\tau^\star-(z-z_0)\Om_\alpha^\star-x_{0\alpha}\Om_z^\star\right\}\partial_\alpha\delta_0,\varphi_z\gg,
\end{eqnarray}
where $\delta_0$ is the $2D$ Dirac measure located at $0$ and $\varphi_z(x_\alpha):=\varphi(x_\alpha,z)$, while symbol $\ll\cdot,\cdot\gg$ denotes the $2D$ distribution by test-function product.
\item When $k=\kappa$, Definition \ref{burgtens} shows that
\begin{eqnarray}
\displaystyle\overline\partial_\beta b^\star_z:=\mathcal{E}^\star_{\beta z}+\epsilon_{\gamma\tau}(x_\gamma-x_{0\gamma})\overline\partial_\beta\omega^\star_\tau\nonumber,
\end{eqnarray}
from which, after multiplication by $\epsilon_{\kappa\alpha}$, it results that:
\begin{eqnarray}
&x_\alpha\overline\partial_\tau\omega_\kappa^\star=-\epsilon_{\kappa\alpha}\overline\partial_\tau b_z^\star
+\epsilon_{\kappa\alpha}\mathcal{E}_{\tau z}^\star+x_{0\alpha}\overline\partial_\tau\omega_\kappa^\star
+(x_\kappa-x_{0\kappa})\overline\partial_\tau\omega_\alpha^\star.\nonumber
\end{eqnarray}
Then, by Lemma \ref{fleurs} with a permutation of indices $\kappa$ and $\alpha$, Eq. (\ref{ombre}) also writes as
\begin{eqnarray}\label{pik'}
\Pi'_\kappa&=&\partial_\alpha\varphi(\hat x)\int_{C_{\epsilon,z}}\left(-\epsilon_{\kappa \alpha}\overline\partial_\beta b_z^\star+\epsilon_{\kappa\alpha}\mathcal{E}_{\beta z}^\star + x_{0\alpha}\overline\partial_\beta\omega_\kappa^\star
-x_{0\kappa}\overline\partial_\beta\omega_\alpha^\star
\right)dx_\beta\nonumber\\
&&\hspace{30pt}+o(1).\nonumber
\end{eqnarray}
On the other hand, from Eq. (\ref{ombre}) and Lemma \ref{fleurs} (i.e. from strain compatibility) it follows that:
\begin{eqnarray}
\Pi'_\kappa&=&
\partial_\alpha\varphi(\hat x)\int_{C_{\epsilon,z}}\left(-\epsilon_{\kappa\beta}
\mathcal{E}_{\beta z}^\star dx_\alpha+\epsilon_{\alpha\beta}
\mathcal{E}_{\beta z}^\star dx_\kappa\right)+o(1)\nonumber\\&=&\partial_\alpha\varphi(\hat x)\int_{C_{\epsilon,z}}\epsilon_{\alpha\kappa}
\mathcal{E}_{\beta z}^\star dx_\beta+o(1).\label{chiho}
\end{eqnarray}
By summing this latter expression of $\Pi'_\kappa$ with Eq. (\ref{chiho}), from the definitions of the Frank and Burgers vector it follows that
\begin{eqnarray}\label{bruno}
\Pi'_\kappa=\frac{1}{2}\partial_\alpha\varphi(\hat x)\epsilon_{\alpha\kappa}
\left(B_z^\star-\epsilon_{\gamma\beta}\Om_\gamma^\star x_{0\beta}\right)+o(1).
\end{eqnarray}
Hence, in the limit $\epsilon\to 0^+$ Eq. (\ref{bruno}) writes as
\begin{eqnarray}\label{foug�es}
\lim_{\epsilon\to 0^+}\Pi'_\kappa=\ll\left\{\frac{1}{2}\epsilon_{\kappa\alpha}B_z^{\star}
-\frac{1}{2}\epsilon_{\kappa\alpha}\epsilon_{\gamma\beta}\Om_\gamma^\star x_{0\beta}\right\}\partial_\alpha\delta_0,\varphi_z\gg.
\end{eqnarray}
\end{itemize}
Therefore, the result is proved on $\Om_z^0$, since
\begin{eqnarray}\label{buissons}
\lim_{\epsilon\to 0^+}\Pi_k(z,\varphi,\epsilon)=\lim_{\epsilon\to 0^+}\Pi'_k(z,\varphi,\epsilon)+\ll\Om^\star_k\delta_0,\varphi_z\gg.
\end{eqnarray}
\newline
$\blacktriangle$ As suggested by Eq. (\ref{marron}), to obtain the result for the entire domain $\Om$ it suffices to integrate Eqs. (\ref{mousse}) and (\ref{bruno}) and expression $\Om^\star_k\varphi(\hat x)$ over $L$, in order to replace $\delta_0$ by the line measure $\delta_L$ in Eqs. (\ref{feuilles}), (\ref{foug�es}) and (\ref{buissons}). By Eqs. (\ref{marron}), (\ref{feuilles}), (\ref{foug�es}) and  (\ref{buissons}), the proof is achieved.{\hfill $\square$}

\subsection{Applications of the main result}\label{appl}
Throughout this section $(x,y,z)$ denotes a generic point of $\Om_L$ and all tensors are written in matrix form in the Cartesian base $(e_x,e_y,e_z)$.
\begin{itemize}
\item \textit{Screw disclocation.} Since $B^\star_\gamma=\Om^\star_z=0$, Eq. (\ref{final2}) yields
\begin{eqnarray}
\displaystyle[\eta^\star_k]=\frac{B^\star_z}{2}\left[\begin{array}{ccc} \partial_y\delta_L \\-\partial_x\delta_L \\ 0\end{array}\right].\nonumber
\end{eqnarray}
This result is easily verified with use of Eq. (\ref{FP2}). One needs to compute $\displaystyle \int_\Om\epsilon_{kpn}\epsilon_{\alpha\beta}\mathcal{E}^\star_{\beta n}\partial_p\partial_\alpha\varphi dV$,
that is to calculate the integral of
\begin{eqnarray}
\frac{B^\star_z}{4\pi}\left[\begin{array}{ccc} \partial_y\partial_x\varphi\frac{\cos\theta}{r}+\partial^2_y\varphi\frac{\sin\theta}{r}
\\  -\partial^2_x\varphi\frac{\cos\theta}{r}-\partial_x\partial_y\varphi\frac{\sin\theta}{r}
\\0\end{array} \right].\nonumber
\end{eqnarray}
By integration by parts, using Gauss-Green's theorem on $\Om$, and recalling that test-functions have compact supports and that $\displaystyle\partial_m\log r=\frac{x_m}{r^2}$, these integrals become
\begin{eqnarray}
-\frac{B^\star_z}{4\pi}\displaystyle\int_\Om
\left[
\begin{array}{c}
\partial_y\varphi\left(\partial_x\frac{\cos\theta}{r}
+\partial_y\frac{\sin\theta}{r} \right)\\ -\partial_x\varphi\left(\partial_x\frac{\cos\theta}{r}+\partial_y\frac{\sin\theta}{r}\right)\\
0
\end{array}
\right]dV
\frac{B^\star_z}{4\pi}\displaystyle\int_\Om
\left[
\begin{array}{c} -\partial_y\varphi\partial_m^2\log r\\
\partial_x\varphi\partial_m^2\log r\\
0
\end{array}
\right]dV.\nonumber
\end{eqnarray}
Hence, from $\Delta\left(\log r\right)=2\pi\delta_L$, the first statement is  verified.
\item \textit{Edge dislocation.} Whereas $\overline\partial_m\omega^\star_k$ identically vanishes on $\Om_L$,  it is easily seen that Eqs. (\ref{final1}) and (\ref{final2}) with $B^\star_z=\Om^\star_z=0$ yield
\begin{eqnarray}
[\eta^\star_k]=B^\star_y\left[\begin{array}{ccc} 0\\0 \\ \partial_x\delta_L
\end{array}\right].\nonumber
\end{eqnarray}
We must compute $\displaystyle [\eta^\star_k]=\int_\Om\epsilon_{pnk}\epsilon_{\alpha\beta}\mathcal{E}^\star_{\beta n}\partial_p\partial_\alpha\varphi dV$.
For $k=1$ and $2$ and with $n\not =3$, the tensor $\epsilon_{\alpha\beta}\mathcal{E}^\star_{\beta n}\partial_p\partial_\alpha\varphi$ equals $\mathcal{E}^{\star}_{yx}\partial_z\partial_y\varphi-\mathcal{E}^{\star}_{yy}\partial_z\partial_x\varphi$ and
$\mathcal{E}^{\star}_{xy}\partial_z\partial_x\varphi-\mathcal{E}^{\star}_{xx}\partial_z\partial_y\varphi$ respectively. By integration by parts, the related integrals vanish. For $k=3$, the integrand is
\begin{eqnarray}
\epsilon_{pnz}\epsilon_{\alpha\beta}\mathcal{E}^\star_{\beta n}\partial_p\partial_\alpha\varphi\mathcal{E}^\star_{xx}\partial_y\partial_y\varphi
+\mathcal{E}^\star_{yy}\partial_x\partial_x\varphi
-2\mathcal{E}^\star_{xy}\partial_y\partial_x\varphi.\nonumber
\end{eqnarray}
Integration by parts provides the expression $\displaystyle\int_\Om-\frac{B_y}{2\pi}\partial_x\varphi \Delta(\log r)dV$, achieving the second verification. 
\newline
\item \textit{Wedge disclination.} Incompatibility reads
\begin{eqnarray}
[\eta^\star_k]=\Om^\star_z\left[\begin{array}{ccc} 0\\0 \\ \delta_L\end{array}\right].\nonumber
\end{eqnarray}
We must calculate $<\eta_k^\star,\varphi>$. For $k=1$, $k=2$, $n\not =0$ and $p=3$ the integrand vanishes. For $k=3$, one computes
\begin{eqnarray}
\epsilon_{pn}\epsilon_{lm}\mathcal{E}^{\star}_{mn}\partial_p\partial_l\varphi&=&
\frac{\Om^\star_z(1-\nu^\star)}{4\pi}\varphi\Delta(\log\frac{r}{R})
+\frac{\Om^\star_{z}(1-\nu^\star)}{4\pi}\varphi\Delta(\log\frac{r}{R})\nonumber\\
&=&\frac{\Om^\star_z}{4\pi}(2*2\pi\delta_L),\nonumber
\end{eqnarray}
achieving the third verification.
\end{itemize}
\section{Distributional analysis of incompatibility for a set of isolated dislocations}\label{distrset}
In the previous section, a single defect line was considered. However, to address the macro-scale physics, homogenisation must be performed on a set of dislocation lines whose number tends to infinity in order to define regular defect density tensors. Therefore, our goal is to introduce appropriate hypotheses that can easily be applied to a set of defect lines and to a regular defect density as well.
\subsection{Governing assumptions for the strain and Frank tensors}\label{assumpt}
Besides the strain Assumptions \ref{asstrain} two measure hypotheses on the strain derivatives are introduced  to replace the local Assumption \ref{asstrainloc} in order to validate Theorem \ref{mainresultloc} in a global framework.
\begin{Ass}\label{assfranktens} 
The strain divergence and trace gradient  $\partial_\alpha\strain_{\alpha i}$ and $\partial_\gamma\strain_{\kappa\kappa}$
are finite Radon measures on $\Om$.\footnote{A finite Radon measure on $\Om$ is a measure bounded on compact subsets of $\Om$.
}
\end{Ass}
\begin{Rem}
No assumption could be made on the complete Lebesgue integrable strain gradient without contradicting the $2D$ examples of Appendix \ref{2Dsection}. On the other hand, it can be shown that the sharp Assumption \ref{assfranktens} are required to demonstrate Proposition \ref{straindecomp}.
\end{Rem}
\begin{Rem}
Assumption \ref{assfranktens} is natural in infinitesimal elasticity if one considers the strain-stress constitutive law and the equilibrium laws. As a consequence, the stress divergence must be a measure on $\Om$.
\end{Rem}
The following Lemmas are needed for the proof of Proposition \ref{straindecomp}.
\begin{lemma}\label{helmholtz}
\begin{itemize}
\item A solenoidal distributional vector field $a_\alpha$ on $\Om_z$ writes as
\begin{eqnarray}\label{firststatement}
a_\alpha=\epsilon_{\alpha\gamma}\partial_\gamma\phi,
\end{eqnarray}
with $\phi\in\mathcal{D}'(\Om_z)$.
\item A symmetric solenoidal distribution tensor $a_{\alpha\beta}$ on $\Om_z$ writes as
\begin{eqnarray}\label{secondstatement}
a_{\alpha\beta}=\epsilon_{\alpha\gamma}\epsilon_{\beta\tau}\partial_\gamma\partial_\tau\psi,
\end{eqnarray} 
with $\psi\in\mathcal{D}'(\Om_z)$.
\end{itemize}
\end{lemma}
{\bf Proof.} 
\begin{itemize}
\item \textit{First statement}.
Let $\phi_0$ be any $x_2$-primitive distribution of $a_1$ (Schwartz, 1957). Then $\partial_2\phi_0=a_1$ and, from the solenoidal property of $a_\alpha$, there exists a distribution $G(x_1)$ s.t. $\partial_1\phi_0+a_2=G(x_1)$. By $x_1$-primitivation of $G(x_1)$, it is easy to find $F(x_1)$ s.t. $\partial_1 F=G(x_1)$, and to verify that $\phi=\phi_0+F(x_1)$ solves the problem.
\newline
\item \textit{Second statement.}
>From $\partial_\alpha a_{\alpha\beta}=0$, there is a distribution $\phi_\beta$ s.t. $a_{\alpha\beta}=\epsilon_{\alpha\gamma}\partial_\gamma\phi_\beta$. Then
$\epsilon_{\alpha\gamma}\partial_\gamma\left(\partial_\beta\phi_\beta\right)\partial_\beta a_{\alpha\beta}=0$ and hence $\partial_\beta\phi_\beta$ is a constant $C$ or equivalently $\partial_\beta(\phi_\beta-\frac{1}{2}Cx_\beta)=0$. From Eq. (\ref{firststatement}), there exists a distribution $\psi$ such that $\phi_\beta-\frac{1}{2}Cx_\beta=\epsilon_{\beta\tau}\partial_\tau\psi$, and hence $a_{\alpha\beta}=\epsilon_{\alpha\gamma}\epsilon_{\beta\tau}\partial_\gamma\partial_\tau\psi+\frac{1}{2}\epsilon_{\alpha\beta}C$. The symmetry of $a_{\alpha\beta}$ implies that $C=0$. {\hfill $\square$}
\end{itemize}
\begin{lemma}\label{divf}
\begin{itemize}
\item For a given $L^1(\Om_z)$-scalar function $f$, there exists an irrotational distribution field $g_\beta$ such that
\begin{eqnarray}
\partial_\beta g_\beta=f\label{divg}.
\end{eqnarray}
\item For a given $L^1$($\Om_z$)-vector function $f_\beta$ such that $\partial_\beta f_\beta=\Delta g$ where $g$ is a $L^1(\Om_z)$ function, there exists a symmetric compatible tensor $g_{\alpha\beta}$ on $\Om_z$ such that
\begin{eqnarray}\label{fbeta}
\partial_\alpha g_{\alpha\beta}=f_\beta.
\end{eqnarray}
\end{itemize}
\end{lemma}
{\bf Proof.} 
\begin{itemize}
\item \textit{First statement.}
It is sufficient to consider an ultra-weak solution (Brezis, 1983) of $\Delta H=f$
and to define  $g_\beta=\partial_\beta H$.
\newline
\item \textit{Second statement.}
By primitivation, there is a non-compatible $L^1(\Om_z)$-field $g^\star_{\alpha\beta}$ such that
$f_1=\partial_1g^\star_{11}, f_2=\partial_2g^\star_{22}$ and $0=g^\star_{21}=g^\star_{12}$. A necessary condition for $g_{\alpha\beta}$ to exist is that $\hat{g}_{\alpha\beta}=g_{\alpha\beta}-g^\star_{\alpha\beta}$ verifies $\partial_\alpha\hat{g}_{\alpha\beta}=0$, or by Lemma \ref{helmholtz} that $\hat{g}_{\alpha\beta}=\epsilon_{\alpha\gamma}\epsilon_{\beta\tau}\partial_\gamma\partial_\tau\phi$ for some gauge distribution $\phi$. In order that $g_{\alpha\beta}$ be compatible on $\Om_z$, $\phi$ must satisfy the following equation, equivalent to the $2D$ compatibility of $g_{\alpha\beta}$ on $\Om_z$:
\begin{eqnarray}
\Delta\Delta\phi=\Delta g^\star_{\kappa\kappa}-\partial_\beta\partial_\alpha g_{\alpha\beta}=\Delta(g^\star_{\kappa\kappa}-g).\label{labo1}
\end{eqnarray}
Up to a harmonic and hence smooth function on $\Om_z$, the solution of Eq. (\ref{labo1}) is the solution of
$\Delta\phi=g^\star_{\kappa\kappa}-g$. Since the right-hand side is $L^1(\Om_z)$, a solution $\phi$ exists in the ultra-weak sense and hence the existence of a symmetric compatible distribution field $g_{\alpha\beta}$ on $\Om_z$ verifying Eq. (\ref{fbeta}) follows.{\hfill $\square$}
\end{itemize}

\begin{lemma}\label{decomp_delta}
For constant $C$ and $C_\beta$, there are a vector $g_\kappa$ and a symmetric, compatible tensor $G_{\alpha\beta}$ on $\Om_z$ such that
\begin{eqnarray}
\partial_\kappa g_\kappa&=&C\delta_0,\label{phibeta1}\\
\partial_\alpha G_{\alpha\beta}&=&C_\beta\delta_0.\label{phibeta}
\end{eqnarray}
\end{lemma}
{\bf Proof.} 
The solutions are given by $g_\kappa=(2\pi)^{-1}\partial_\kappa\log r$
and $\displaystyle G_{\alpha\beta}=\frac{1}{2}\left(\partial_\alpha H_\beta\right.$ $\left.+\partial_\beta H_\alpha\right)$, where
\begin{eqnarray}\label{existenceG}
H_1&=&\frac{C_1}{2\pi}\left(\frac{3}{2}\log r-\frac{x_1^2}{2r^2}\right)-C_2 \frac{x_1x_2}{4\pi r^2},\nonumber\\
H_2&=&\frac{C_2}{2\pi}\left(\frac{3}{2}\log r-\frac{x_2^2}{2r^2}\right)-C_1 \frac{x_1x_2}{4\pi r^2}\nonumber.
\end{eqnarray}{\hfill $\square$}
\begin{lemma}\label{straindecomp}
Under Assumptions \ref{asstrain} and \ref{assfranktens}, the strain components can be put in the form:
\begin{eqnarray}
\strain_{\kappa z}&=&E_\kappa+e_\kappa,\label{straindecomp1}\\
\strain_{\alpha\beta}&=&E_{\alpha\beta}+e_{\alpha\beta},\label{straindecomp2}
\end{eqnarray}
where vector $E_\kappa$ has a vanishing curl on $\Om_z$ for any given $z$ while vector $e_\kappa$ is $o(r^{-2})$ as $r\to 0^+$, and 
where tensor $E_{\alpha\beta}$ is compatible on $\Om_z$ for any given $z$ while tensor $e_{\alpha\beta}$ is $o(r^{-2})$ as $r\to 0^+$.
\end{lemma}
{\bf Proof.} 
By Assumption \ref{assfranktens}, $\partial_\kappa\strain_{\kappa i}$ is a Radon measure on $\Om_z$, and hence writes by Radon-Nykod\'ym's decomposition theorem as
\begin{eqnarray}\label{police2}
\partial_\kappa\strain_{\kappa i}=f_ i+\phi_i,\label{kappabeta}
\end{eqnarray}
where $f_i\in L^1(\Om_z)$ and where $\phi_i$ is a Radon measure on $\Om_z$ singular with respect to Lebesgue's measure. As a mere consequence of the smoothness of $\partial_\kappa\strain_{\kappa i}$ on $\Om_z^0$, $\phi_i$ is a concentrated measure on $\Om_z$ and hence is proportional to the Dirac mass $\delta_0$,
\begin{eqnarray}\label{phii}
\phi_i=C_i\delta_0=(2\pi)^{-1}C_i\partial_\kappa^2\log r.
\end{eqnarray}
$\blacktriangle$ \textit{First statement.}
\begin{itemize}
\item By Eqs. (\ref{police2}), (\ref{phii}) with $i=z$, and Lemma \ref{divf}, there exists an irrotational $g_\kappa$ such that
\begin{eqnarray}
\partial_\kappa\left(\strain_{\kappa z}-g_\kappa-(2\pi)^{-1}C_z\partial_\kappa\log r\right)=0,\nonumber
\end{eqnarray}
in such a way that, by Lemma \ref{helmholtz},
\begin{eqnarray}
\strain_{\kappa z}-g_\kappa-(2\pi)^{-1}C_z\partial_\kappa\log r=\epsilon_{\kappa\gamma}\partial_\gamma\psi,\label{decomp}
\end{eqnarray}
where $\psi$ is a distribution. Apply the curl operator to Eq. (\ref{decomp}) and take into account the irrotational property of $g_\kappa$ in such a way that $\Delta\psi=\epsilon_{\kappa\beta}\partial_\beta\strain_{\kappa z}$.
Since $\strain_{\kappa z}$ is a $L^1$-vector, its curl is a first-order distribution\footnote{Following Schwartz (1957), a distribution is of order $1$ if it defines a linear continuous map on $\mathcal{C}^1_c(\Om)$.} and hence, by the strain compatibility which ensures the curl of $\strain_{\kappa z}$ to be a constant $K$ on $\Om_z^0$ and a combination of the Dirac mass and its first-order derivatives at the origin (Schwartz, 1957), writes as $K+c\delta+c_\gamma\partial_\gamma\delta$. \newline
\item Now in the resulting equation
\begin{eqnarray}
\partial_\beta\left(\epsilon_{\kappa\beta}\strain_{\kappa z}-(2\pi)^{-1}c\partial_\beta\log r-\frac{K}{2}x_\beta\right)=c_\gamma\partial_\gamma\delta,\label{resulteq}
\end{eqnarray}
the term on the left-hand side is the divergence of a $L^1$-vector, and hence Eq. (\ref{resulteq}) has no distributional solution unless $c_\gamma=0$.\newline
\item It results that $\Delta\psi=K+(2\pi)^{-1}c\Delta(\log r)$
provides a gauge field $\psi$ which writes as
\begin{eqnarray}\label{psi}
\psi=h+(2\pi)^{-1}c\log r,
\end{eqnarray}
where $h$ is a solution of $\Delta h=K$ on $\Om_z$. It is easily verified that the curl of $\psi$ is $o(r^{-2})$ as $r\to 0^+$.\newline
\item Defining $E_\kappa=g_\kappa+(2\pi)^{-1}C_z\partial_\kappa\log r$ and $e_\kappa=\epsilon_{\kappa\gamma}\partial_\gamma\psi$ in Eq. (\ref{decomp}) achieves the first statement proof.
\end{itemize}
$\blacktriangle$ \textit{Second statement.}
\begin{itemize}
\item Let us prove that the divergence of $f_i$ is the Laplacian of an $L^1(\Om_z)$ function. In fact, since $\eta^\star_z$ writes as
\begin{eqnarray}
\eta_z^\star=\partial_\alpha\left(\partial_\alpha\mathcal{E}_{\kappa\kappa}^\star-\partial_\beta\strain_{\alpha\beta}\right),\label{etaz}
\end{eqnarray}
it is from Assumption \ref{assfranktens} a concentrated first-order distribution writing as a combination of the Dirac mass and its first-order derivatives. Hence:
\begin{eqnarray}
\partial_\beta f_\beta&=&\partial_\alpha\partial_\beta\strain_{\alpha\beta}-\partial_\beta\phi_\beta=\Delta\strain_{\kappa\kappa}-\eta^\star_z-\partial_\beta\phi_\beta=\Delta\strain_{\kappa\kappa}-\hat c\delta_0-\hat c_\gamma\partial_\gamma\delta_0\nonumber\\
&=&\Delta\left(\strain_{\kappa\kappa}-\overline c\log r-\overline c_\gamma\partial_\gamma\log r\right),
\end{eqnarray}
where $\hat c, \hat c_\gamma, \overline c, \overline c_\gamma$ are constants.
\item From Eqs. (\ref{police2}), (\ref{phibeta}), (\ref{fbeta}) and Lemma \ref{divf}, there exists a compatible $g_{\kappa\beta}$ such that
\begin{eqnarray}
\partial_\kappa\left(\strain_{\kappa\beta}-g_{\kappa\beta}-G_{\kappa\beta}\right)=0,
\end{eqnarray}
in such a way that, by Lemma \ref{helmholtz},
\begin{eqnarray}\label{decompstrain}
\strain_{\kappa\beta}-g_{\kappa\beta}-G_{\kappa\beta}=\epsilon_{\kappa\gamma}\epsilon_{\beta\tau}\partial_\gamma\partial_\tau A,
\end{eqnarray}
for some gauge field $A\in\mathcal{D}'(\Om_z)$ verifying, by the compatibility of $g_{\kappa\beta}$ and $G_{\kappa\beta}$ on $\Om_z$, the relation
\begin{eqnarray}
\eta^\star_z=\Delta\Delta A \quad\mbox{on}\quad\Om_z.\label{doublelapl}
\end{eqnarray}
Hence, since the left-hand side writes as a combination
of derivatives of $\delta_0$ of order lower or equal to $1$, the field $A$ is the solution of
$\Delta A =\left(a+a_\gamma\partial_\gamma\right)\log r$ with constant $a,a_\gamma$, up to a smooth harmonic function on $\Om_z$. It follows that $A =\left(a+a_\gamma\partial_\gamma\right)\left(\frac{r^2}{4}(\log r-1)\right)$ is a $\mathcal{C}^0(\Om_z)$ solution  of Eq. (\ref{doublelapl}) such that:
\begin{eqnarray}\label{Aor2}
\partial_\kappa\partial_\beta A \quad\mbox{is}\quad o(r^{-2})\quad\mbox{as}\quad r\to 0^+.
\end{eqnarray}
\item The proof is complete with the definitions $E_{\kappa\beta}=G_{\kappa\beta}+g_{\kappa\beta}$ and $e_{\kappa\beta}=\epsilon_{\kappa\gamma}\epsilon_{\beta\tau}\partial_\gamma\partial_\tau A$ in Eqs. (\ref{decompstrain}) and (\ref{Aor2}).{\hfill $\square$}
\end{itemize}
\subsection{Mesoscopic incompatibility for a set of isolated defect lines}\label{incset}
\begin{theorem}\label{mainresultglob}[Main $2D$ result]
Under Assumptions \ref{asstrain} and \ref{assfranktens}, for a set $\mathcal{L}$ of isolated dislocations parallel to the $z$-axis and located at the positions $x_\beta^L $, $L\in\mathcal{L}$, incompatibility as defined by Eq. (\ref{eta-om}) is the vectorial first order distribution
\begin{eqnarray}
\eta_k^{\star}=\delta_{kz}\eta_z^{\star}+\delta_{k\kappa}\eta_\kappa^{\star},\label{eta_kglob}
\end{eqnarray}
where
\begin{itemize}
\item its vertical component is
\begin{eqnarray}
\eta^\star_z=\displaystyle\sum_{L\in\mathcal{L}}\left(\Om^\star_z\delta_L+\epsilon_{\alpha\gamma}\left(B^\star_\gamma+\epsilon_{\beta\gamma}(x_\beta^L-x_{0\beta})\Om^\star_z\right)\partial_\alpha\delta_L\right),\label{final1glob}
\end{eqnarray}
\item its planar components are
\begin{eqnarray}
\eta^\star_\kappa=\displaystyle\sum_{L\in\mathcal{L}}\frac{1}{2}\epsilon_{\kappa\alpha}B^\star_z\partial_\alpha\delta_L.\label{final2glob}
\end{eqnarray}
\end{itemize}
\end{theorem}
{\bf Proof.} 
>From Lemma \ref{straindecomp} the strain $\strain_{\beta n}$ ($n=\alpha$ or $z$ ) is decomposed in compatible parts ($E_\beta$ and $E_{\alpha\beta}$) and $o(r^{-2})$ parts ($e_\beta$ and $e_{\alpha\beta}$) to which the demonstration may be limited by linearity. Since from Eqs. (\ref{psi}) and (\ref{Aor2}) the gradients $\partial_\gamma e_\beta,\partial_\gamma e_{\alpha\beta}$ are $o(r^{-3})$ for $r\to 0^+$, the proof of Theorem \ref{mainresultloc} can be followed for every $L\in \mathcal{L}$ as soon as $\strain_{\beta z}$ is replaced by $e_\beta$ and $\strain_{\beta\tau}$ by $e_{\beta\tau}$. However, since the dislocations are located at positions $x_\beta^L$ instead of $0$, an additional shift $x_\beta^L$ is required in Eq. (\ref{final1glob}).{\hfill $\square$}

\subsection{Mesoscopic defect densities in $2D$ incompatible elasticity}\label{defectdens}
Since the tensors $\Theta^\star_{ik},\Lambda^\star_{ik},\alpha^\star_{ik}$ vanish for $i\not=z$, the $2D$ densities for an ensemble $\mathcal{L}$ of rectilinear dislocations write as follows\footnote{Various notations are used in the literature to represent the defect densities. In particular, Nye (1953), Kr\"oner (1980) and Kleinert (1989) give different definitions of the dislocation density and contortion tensors (without considering disclinations in the first two cases). We here follow Kr\"oner's and Kleinert's notations for $\alpha^\star_{ij}$ and Nye's original definition of $\kappa^\star_{ij}$, with Nye's $\alpha^\star_{ij}$ here denoted by $\alpha^\star_{ji}$. It should be recalled that the term "contortion" was introduced by Kondo (1952).}:
\begin{Def}
\begin{eqnarray}
\Theta^\star_k&:=&\displaystyle\sum_{L\in\mathcal{L}}\delta_{kz}\Om^{\star L}_z\delta_L\label{theta2D}\\
\Lambda^\star_k&:=&\displaystyle\sum_{L\in\mathcal{L}}B^{\star L}_k\delta_L\label{lambda2D},\\
\alpha^\star_k&:=&\alpha^\star_{zk}=\Lambda^\star_k-\delta_{k\alpha}\epsilon_{\alpha\beta}\Theta^\star_z(x_\beta-x_{0\beta})\label{alpha2D}.
\end{eqnarray}
\end{Def}
Moreover, in the $2D$ case, the contortion tensor writes as:
\begin{eqnarray}\label{kappa2D}
\kappa^\star_{ij}=\delta_{iz}\alpha^\star_j-\frac{1}{2}\alpha^\star_z\delta_{ij}.
\end{eqnarray}
The following result expresses the incompatibility in terms of $\kappa_{ij}^\star$:
\begin{theorem}\label{kronerident}
Under Assumptions \ref{asstrain} and \ref{assfranktens}, the mesoscopic strain incompatibility for a set $\mathcal{L}$ of rectilinear dislocations writes as
\begin{eqnarray}
\eta^\star_k&=&\Theta^\star_k+\epsilon_{\alpha\beta}\partial_\alpha\kappa_{k\beta}^\star,\label{kroner1}
\end{eqnarray}
or equivalently as $\eta^\star_k=\Theta^\star_k+\epsilon_{k\alpha l}\partial_\alpha\kappa^\star_{zl}$.{\hfill $\square$}
\end{theorem}
{\bf Proof.} 
Consider any straight dislocation $L\in\mathcal{L}$ located at a given $x^L\in\Om$. From Theorem \ref{mainresultglob}, incompatibility writes as
\begin{eqnarray}\label{translat}
\eta^\star_k&=&\delta_{kz}\left(\Om^\star_z\delta_L+\epsilon_{\alpha\gamma}\left(B^\star_\gamma+\epsilon_{\beta\gamma}(\hat x_\beta-x_{0\beta})\Om^\star_z\right)\partial_\alpha\delta_L\right)\nonumber\\
&+&\delta_{k\kappa}\frac{1}{2}\epsilon_{\kappa\alpha}B^\star_z\partial_\alpha\delta_L.
\end{eqnarray}
Taking into account Eqs. (\ref{theta2D}), (\ref{lambda2D}), (\ref{alpha2D}), and (\ref{kappa2D}), and the relation
\begin{eqnarray}
\partial_\alpha\left((x_\beta-x_{0\beta})\delta_L\right)=\partial_\alpha\left((x_\beta^L-x_{0\beta})\delta_L\right)=(x_\beta^L-x_{0\beta})\partial_\alpha\delta_L,\nonumber
\end{eqnarray}
it results from Theorem \ref{mainresultglob} that incompatibility can be written in the alternative formulation
\begin{eqnarray}
\eta^\star_k(x^L)&=&\Theta^\star_k(x^L)+\epsilon_{\alpha\beta}\partial_\alpha\kappa_{k\beta}^\star(x^L),
\end{eqnarray}
or equivalently as $\eta^\star_k(x^L)=\Theta^\star_k(x^L)+\epsilon_{k\alpha l}\partial_\alpha\kappa^\star_{zl}(x^L)\nonumber$.
The result follows after summation on $L\in\mathcal{L}$ and using Eqs. (\ref{theta2D}), (\ref{lambda2D}), (\ref{alpha2D}), and (\ref{kappa2D}).
{\hfill $\square$}
\newline\newline
First of all, the tensor $\overline\partial_j\overline\partial_l u^\star_k$ is defined on the entire $\Om$ in a similar way as $\overline\partial_j\omega^\star_k$:
\begin{Def}
\begin{eqnarray}
\overline\partial_j\overline\partial_l u^\star_k:=\partial_j\strain_{kl}+\epsilon_{kpl}\overline\partial_j\omega^\star_p.
\end{eqnarray}
\end{Def}
By Proposition \ref{displ}, the displacement field $u^\star_k$ is a multivalued function of index $2$, which is obtained on $\Om_L$ by recursive line integration of $\partial_j^{(s)}\partial_l^{(s)} u^\star_k=\partial_j^{(s)}\left(\strain_{kl}+\omega_{kl}^\star\right)$ and hence by recursive integration of $\overline\partial_j\overline\partial_l u^\star_k$. 
\begin{Rem}
In the situation where, for a particular selection of the reference point, the dislocations have vanishing Burgers vectors, the disclination density equals the incompatibility
\begin{eqnarray}
\epsilon_{\alpha\beta}\partial_\alpha\overline\partial_\beta\omega_k^\star=\Theta_k^\star=\eta^\star_k.
\end{eqnarray}
Using an arbitrary reference point, this expression is certainly false in the general case where disclinations coexist with dislocations. 
Moreover, the tensor $\overline\partial_j\overline\partial_l u^\star_k$ does not provide relevant information in terms of defect densities since $\epsilon_{ijl}\overline\partial_j\overline\partial_l u^\star_k=0$ on $\Om$. 
\end{Rem}
The mesoscopic vectors and tensors $\Theta^\star_k, \Lambda^\star_k, \alpha^\star_k,\kappa^\star_k$ and $\eta^\star_k$ are concentrated distributions on the defect lines which provide all the information on dislocation and disclination densities. However, homogenisation to the macro-scale still requires to clarify their link with the multiple-valued rotation and displacement fields. 
In order to resolve this problem, the tensors $\overline\partial_j\omega^\star_k$ and $\overline\partial_j\overline\partial_l u^\star_k$ are completed by appropriate concentrated effects in the defect lines, without however modifying their relationship with the multiple-valued displacement and rotation fields defined in $\Om_L$.
\begin{Def}\label{defmeso}
\begin{eqnarray}
\eth_\beta\omega_k^\star&:=&\overline\partial_\beta\omega^\star_k-\kappa^\star_{k\beta},\label{disclindens}\\
\eth_\alpha\eth_\beta u^\star_k&:=&\overline\partial_\alpha\overline\partial_\beta u^\star_k-\epsilon_{kp\beta}\kappa^\star_{p\alpha}=\partial_\alpha\strain_{k\beta}+\epsilon_{kp\beta}\eth_\alpha\omega^\star_p.\label{dislocdens}
\end{eqnarray}
\end{Def}
\begin{theorem}\label{nouveautenseur}
The vector and tensor distributions $\eth_\beta\omega_k^\star$ and $\eth_\alpha\eth_\beta u^\star_k$ verify:
\begin{eqnarray}
&&\hspace{-55pt}\mbox{\scriptsize{MESOSCOPIC DISCLINATION DENSITY}}\hspace{40pt} \Theta^\star_k=\epsilon_{\alpha\beta}\partial_\alpha\eth_\beta\omega_k^\star,\label{disclindens2}\\
&&\hspace{-55pt}\mbox{\scriptsize{MESOSCOPIC DISLOCATION DENSITY}}\hspace{40pt} \alpha^\star_k=\epsilon_{\alpha\beta}\eth_\alpha\eth_\beta u^\star_k.\label{dislocdens2}
\end{eqnarray}
\end{theorem}
{\bf Proof.} 
The first statement is a mere consequence of Eq. (\ref{kroner1}) while 
the second one follows from Eq. (\ref{disclindens}) by simple calculations, noting that $\overline\partial_m\omega^\star_m=0$ on $\Om$ and that
\begin{eqnarray}
\alpha^\star_k=\kappa^\star_{zk}-\kappa^\star_{pp}\delta_{zk}.
\end{eqnarray}
{\hfill $\square$}

\begin{Rem}
Eqs. (\ref{disclindens2}) and (\ref{theta2D}) directly show that
\begin{eqnarray}
\int_S\epsilon_{\alpha\beta}\partial_\alpha\eth_\beta \omega_k^\star dS=\int_S\Theta^\star_k dS=\sum_{L\in\mathcal{L}_C}\Om^{\star L}_k,\label{dislocdens3}
\end{eqnarray}
where the domain $S$ is bounded by the counterclockwise-oriented Jordan curve $C$, which encloses once each defect line of the subset $\mathcal{L}_C$ of $\mathcal{L}$.
Similarly, Eqs. (\ref{dislocdens2}) and (\ref{alpha2D}) show that
\begin{eqnarray}
\int_S\epsilon_{\alpha\beta}\eth_\alpha\eth_\beta u_k^\star dS&=&\int_S\left(\Lambda^\star_k-\delta_{k\alpha}\epsilon_{\alpha\beta}\Theta^\star_z(x_\beta-x_{0\beta})\right)dS,\nonumber\\
&=&\sum_{L\in\mathcal{L}_C}\left(B^{\star L}_k-\delta_{k\alpha}\epsilon_{\alpha\beta}\Om^{\star L}_z(x_\beta^L-x_{0\beta})\right).\label{dislocdens3}
\end{eqnarray}
\end{Rem}
\begin{Rem}
The vector $\overline\partial_l\omega_z^\star$ does not verify Stokes theorem, neither in the classical sense, since $\epsilon_{\alpha\beta}\partial_\alpha\overline\partial_\beta\omega_z^\star$ is singular at $x^L$, nor in a measure theoretical sense, since 
$\epsilon_{\alpha\beta}\partial_\alpha\overline\partial_\beta\omega_z^\star$ is not a measure but a first-order distribution given by Eq. (\ref{final1glob}). 
As often observed in the literature, even in an inappropriate context, a formal use of Stokes theorem  may give a correct final result.
We here prefer to avoid any confusion and hence to mention that, in view of a clarification of Stokes' theorem in the context of defective crystals, the following formula holds and can be proved as a consequence of the previous definitions:
\begin{eqnarray}\label{stokesdefects}
\int_C\eth_l\omega_z^\star dx_l=\int_{S_C}\epsilon_{\alpha\beta}\partial_\alpha\eth_l\omega_z^\star dS.
\end{eqnarray}
\end{Rem}
\section{Macroscopic analysis}\label{macroanal}
\subsection{A first approach to homogenisation from meso- to macro-scale}\label{homo}
The mesoscopic results given in the previous sections are now homogenised (in an appropriate manner, whose description (Kr\"oner, 2001) is not the purpose of this paper). Indeed, in the context of linear elasticity, the macroscopic elastic strain $\mathcal{E}_{ij}$ is obtained by averaging the mesoscopic stress $\sigma_{ij}^\star$ and hence the macroscopic elastic incompatibility $\eta_{ik}$ is obtained by averaging the mesoscopic incompatibility $\eta_{ik}^\star$. Moreover the defect densities are homogenised and the macroscopic counterparts of $\Theta_k^\star,\Lambda_k^\star,\alpha_k^\star$ and $\kappa_{ij}^\star$ write as $\Theta_k,\Lambda_k,\alpha_k$, and $\kappa_{ij}$, with
\begin{eqnarray}\label{alpha_macro}
\alpha_k=\kappa_{zk}-\kappa_{pp}\delta_{zk}\quad\mbox{and}\quad\kappa_{ij}=\delta_{iz}\alpha_j-\frac{1}{2}\alpha_z\delta_{ij}.
\end{eqnarray}
\begin{Def}[Macroscopic Frank and Burgers tensors]
The Frank and Burgers vectors crossing a macroscopic surface $S$ are defined as
\begin{eqnarray}
\Om_k(S)&:=&\int_S\Theta_k dS,\label{frankmacro}\\
B_k(S)&:=&\int_S\Lambda_k dS.\label{burgers1macro}
\end{eqnarray}
\end{Def}
By homogenisation of Eqs. (\ref{disclindens}) and (\ref{dislocdens}), the macroscopic counterparts of Definition \ref{defmeso} write as follows:
\begin{Def}\label{defmacro}
\begin{eqnarray}\label{delta_m_macro}
\eth_\beta\omega_k&:=&\epsilon_{kpq}\partial_p\mathcal{E}_{q\beta}-\kappa_{k\beta},\label{optimal}\\
\eth_\alpha\eth_\beta u_k&:=&\partial_\alpha\mathcal{E}_{k\beta}+\epsilon_{kp\beta}\eth_\alpha\omega_p,\label{transportation}
\end{eqnarray}
where $\mathcal{E}_{k\beta}$ and $\kappa_{k\beta}$ define the macroscopic elastic strain and contortion.
\end{Def}
Moreover, the macroscopic counterpart of Theorem \ref{kronerident} (i.e. the fundamental equation "inc $\mathcal{E}$ = $\Theta$ + curl $\kappa$" of the continuum theory of defects by Kr\"oner (1980)\footnote{Note that different sign conventions for the rotation vector and incompatibility apply in Kr\"oner's work.}) and Theorem \ref{nouveautenseur} now follow from homogenisation of the mesoscopic defect densities and from Definition \ref{defmacro}:
\begin{theorem}\label{macrodens}
\begin{eqnarray}
&&\hspace{-40pt}\mbox{\scriptsize{KR\"ONER'S IDENTITY}}\hspace{110pt}\eta_k=\Theta_k+\epsilon_{\alpha\beta}\partial_\alpha\kappa_{k\beta},\label{incomp_macro}\\
&&\hspace{-40pt}\mbox{\scriptsize{MACROSCOPIC DISCLINATION DENSITY}}\hspace{29pt}\Theta_k=\epsilon_{\alpha\beta}\partial_\alpha\eth_\beta\omega_k,\label{disclindensmacro}\\
&&\hspace{-40pt}\mbox{\scriptsize{MACROSCOPIC DISLOCATION DENSITY}}\hspace{33pt}\alpha_k=\epsilon_{\alpha\beta}\eth_\alpha\eth_\beta u_k.\label{dislocdensmacro}
\end{eqnarray}
\end{theorem}
\begin{Rem}\label{stokesmacro}
By Stokes' theorem, if $S$ is  a region enclosed by a curve $C$, which might have only fractal regularity (Harrison and Norton, 1992), then
$\displaystyle\Om_k(S)=\int_C\eth_\beta\omega_k dx_\beta$. Moreover, in the absence of disclinations, $\displaystyle B_k(S)=\int_S\alpha_k dS$.
\end{Rem}
The macroscopic density tensors $\Lambda_k$ and $\kappa_{ij}$, as obtained from the single-valued mesoscopic densities, have a geometrical interpretation (Kr\"oner, 1980; Anthony, 1970) which will be discussed in the following section. Indeed, $\alpha_k$ is directly related to the torsion of a body submitted to an incompatible purely elastic deformation to which a non-Riemannian connexion is attached due to the contortion $\kappa_{ij}$.
\subsection{The non-Riemannian macroscopic body}\label{nonriemannian}
The following geometric objects are introduced after homogenisation of the well-defined mesoscopic elastic strain and defect densities, in order to provide the model of a macroscopic body endowed with a law of parallel displacement together with internal torsion accounting for the defective crystal structure. 
\begin{Def}\label{defgeom}
\begin{eqnarray}
&&\hspace{-28pt}\mbox{\scriptsize{METRIC TENSOR:}}\hspace{41pt}g_{ij}:=\delta_{ij}-2\mathcal{E}_{ij},\label{metric}\\
&&\hspace{-28pt}\mbox{\scriptsize{TORSION:}}\hspace{66pt} T_{k;ij}:=-\frac{1}{2}\epsilon_{pij}\left(\alpha_{pk}-\epsilon_{kmn}\Theta_{pm}(x_n-x_{0n})\right),\label{torsion}\\
&&\hspace{-28pt}\mbox{\scriptsize{SYMMETRIC CHRISTOFFEL SYMBOLS:}}\nonumber\\
&&\hspace{79pt}\tilde\Gamma_{k;ij}:=\frac{1}{2}\left(\partial_i g_{kj}+\partial_j g_{ki}-\partial_k g_{ij}\right),\label{symmcon}\\
&&\hspace{-28pt}\mbox{\scriptsize{CONTORTION:}}\hspace{39pt}\Delta\Gamma_{k;ij}:=T_{j;ik}+T_{i;jk}-T_{k;ji},\label{contortion}\\
&&\hspace{-28pt}\mbox{\scriptsize{NON SYMMETRIC CHRISTOFFEL SYMBOLS:}}\nonumber\\
&&\hspace{79pt}\Gamma_{k;ij}:=\tilde\Gamma_{k;ij}-\Delta\Gamma_{k;ij}.\label{connexion}
\end{eqnarray}
\end{Def}
\begin{Rem}
The metric of the actual configuration $R(t)$ is $\delta_{ij}$. Therefore, as required (cf Introduction and Remark \ref{trippes}) the reference configuration $R_0$ is nowhere used to introduce the above objects. 
\end{Rem}
Since small displacements are considered, no distinction is to be made between upper and lower indices.
\begin{lemma}\label{propgeom}
The tensor $g_{ij}$ defines a Riemannian metric. The symmetric Christoffel symbols $\tilde\Gamma_{k;ij}$ define a symmetric connexion compatible with this metric, while $T_{k;ij}$ and $\Delta\Gamma_{k;ij}$ are skew-symmetric tensors w.r.t. $i$ and $j$ and $i$ and $k$, respectively. Moreover, the components of $T_{k;ij}$ for $i=z$ or $j=z$ vanish in the $2D$ case.
\end{lemma}
{\bf Proof.} 
The first statements follow from basic definitions (Dubrovin et al., 1992; Schouten, 1954) while the last one follows from the fact that, in the $2D$ case, $\alpha_{pk}(x^L)$ and $\Theta_{pm}(x^L)$ for $L\in\mathcal{L}$ are proportional to $\tau_p\delta_L(x^L)$ with $\tau_p$ standing for the tangent vector to the defect line.{\hfill $\square$}
\begin{proposition}\label{gamma}
The Cristoffel symbols $\Gamma_{k;ij}$ define a non-symmetric connexion compatible with the metric $g_{ij}$ and whose torsion writes as $T_{k;ij}$.\footnote{
In the literature, a so-called Bravais' crystal is a macroscopic body endowed with a lattice where parallel displacement along the crystallographic lines is defined by the connexion $\Gamma_{k;ij}$ of Theorem \ref{gamma} and where the metric is not defined by Eq. (\ref{metric}), but by the motion of an internal observer who would measure his own displacement by counting the atomic lattice steps, without feeling the body torsion (Kr\"oner, 1980).}
\end{proposition}
{\bf Proof.} 
It is easy to verify (Dubrovin et al., 1992) that $\Gamma_{k;ij}$ is a connexion since $\tilde\Gamma_{k;ij}$ is a connexion and $\Delta\Gamma_{k;ij}$ is a tensor. Denoting by $\nabla_k$ (resp. $\tilde\nabla_k$) the covariant gradient w.r.t. $\Gamma_{k;ij}$ (resp. $\tilde\Gamma_{k;ij}$), and recalling that a connexion is compatible with the metric $g_{ij}$ if the covariant gradient of $g_{ij}$ w.r.t. this connexion vanishes, we find by Eq. (\ref{connexion})
\begin{eqnarray}
\nabla_{k}g_{ij}:&=&\partial_k g_{ij}-\Gamma_{l;ik}g_{lj}-\Gamma_{l;jk}g_{li}\nonumber\\
&=&\tilde\nabla_{k}g_{ij}+\Delta\Gamma_{l;ik}g_{lj}+\Delta\Gamma_{l;jk}g_{li}\label{tilde},
\end{eqnarray}
where in the right-hand side, the $1^{st}$ term vanishes by Lemma \ref{propgeom} while the $2^{nd}$ and $3^{rd}$ terms cancel each other since $\Delta\Gamma_{l;jk}g_{li}=\Delta\Gamma_{i;jk}=-\Delta\Gamma_{j;ik}$. It results that the connexion torsion, i.e. the skew-symmetric part of $\Delta\Gamma_{j;ik}$ w.r.t. $i$ and $k$, writes as
\begin{eqnarray}
&&\hspace{-30pt}\frac{1}{2}\left(\Delta\Gamma_{j;ik}-\Delta\Gamma_{j;ki}\right)=-\frac{1}{2}\left(\Delta\Gamma_{i;jk}-\Delta\Gamma_{k;ji}\right)=\frac{1}{2}\bigl(\left(\Delta\Gamma_{k;ij}-\Delta\Gamma_{i;kj}\right)+\nonumber\\
&&\hspace{87pt}\left(\Delta\Gamma_{k;ji}-\Delta\Gamma_{k;ij}\right)-\left(\Delta\Gamma_{i;jk}-\Delta\Gamma_{i;kj}\right)\bigr)\label{calcul}.
\end{eqnarray}
Observing that the $1^{st}$ term in the right-hand side of Eq. (\ref{calcul}) writes as $\Delta\Gamma_{k;ij}$ while, by Definition \ref{defgeom} (Eq. (\ref{contortion})), the left-hand side and the two remaining terms of the right-hand side of Eq. (\ref{calcul}) are equal to $T_{j;ik}, T_{k;ji}$ and $-T_{i;jk}$, respectively, the proof is complete.{\hfill $\square$}
\newline\newline
The following result shows $\Delta\Gamma_{k;ij}$ as directly linked to the contortion $\kappa_{ij}$. 
\begin{proposition}\label{deltagamma}
In the $2D$ case, the contortion tensor $\Delta\Gamma_{k;ij}$ writes in terms of $\kappa_{ij}$ as
\begin{eqnarray}
\Delta\Gamma_{k;ij}=\delta_{k\kappa}\left(\delta_{i\alpha}\delta_{j\beta}\epsilon_{\kappa\alpha}\kappa_{z\beta}\right)
+\delta_{i\alpha}\delta_{jz}\epsilon_{\alpha\tau}\kappa_{\tau\kappa}&+&\delta_{iz}\delta_{j\beta}\epsilon_{\beta\tau}\kappa_{\tau\kappa}
\nonumber\\
&-&\delta_{kz}\delta_{i\alpha}\delta_{j\beta}\epsilon_{\alpha\beta}\kappa_{zz}.\nonumber
\end{eqnarray}
\end{proposition}
{\bf Proof.} 
For $k=z$, by Definition \ref{defgeom}, the last statement of Lemma \ref{propgeom}, and Eq. (\ref{alpha_macro}), it is found that $\Delta\Gamma_{z;ij}=\Delta\Gamma_{z;\alpha\beta}\delta_{i\alpha}\delta_{j\beta}$, with
\begin{eqnarray}
\Delta\Gamma_{z;\alpha\beta}=T_{z;\alpha\beta}=-\frac{1}{2}\epsilon_{\alpha\beta}\alpha_z
&=&-\epsilon_{\alpha\beta}\kappa_{zz}\nonumber\\
&=&-\frac{1}{2}\epsilon_{\alpha\tau}\delta_{\tau\beta}\alpha_z=\epsilon_{\alpha\tau}\kappa_{\tau\beta}.\nonumber
\end{eqnarray}
For $k=\kappa$, by Definition \ref{defgeom} and the last statement of Lemma \ref{propgeom}, it is found that
\begin{eqnarray}
\Delta\Gamma_{\kappa;ij}=\delta_{i\alpha}\delta_{j\beta}\left(T_{\kappa;\alpha\beta}+T_{\beta;\alpha\kappa}+T_{\alpha;\beta\kappa}\right)+\delta_{i\alpha}\delta_{jz}T_{z;\alpha\kappa}+\delta_{iz}\delta_{j\beta}T_{z;\beta\kappa},\nonumber
\end{eqnarray}
with $T_{z;\xi\kappa}=\epsilon_{\xi\tau}\kappa_{\tau\kappa}$
and $\displaystyle T_{\xi;\tau\nu}=-\frac{1}{2}\epsilon_{\tau\nu}\left(\alpha_\xi+\epsilon_{\xi\gamma}\Theta_z(x_\gamma-x_{0\gamma})\right)$.
Since the combination of the terms in $\Theta_z$ vanish in $\Delta\Gamma_{\kappa;ij}$, the proof is completed by observing that $\epsilon_{\alpha\beta}\alpha_\kappa+\epsilon_{\kappa\alpha}\alpha_\beta=(\epsilon_{\alpha\kappa}\epsilon_{\tau\nu})\epsilon_{\tau\beta}\alpha_\nu=\epsilon_{\alpha\kappa}\alpha_\beta=\epsilon_{\alpha\kappa}\kappa_{z\beta}$.{\hfill $\square$}
\newline
The following definition introduces two differential forms related, on the one hand (by Definitions \ref{defmacro} and \ref{defgeom}, and Theorem \ref{macrodens} and Proposition \ref{gamma}) to the homogenisation of the well-defined mesoscopic defect measures and, on the other hand, as shown by the forthcoming theorem, to macroscopic incompatible rotation and distortion vectors.
\begin{Def}\label{formdiff}
The following differential forms are introduced:
\begin{eqnarray}
d\omega_j&:=\eth_\beta\omega_jdx_\beta,\label{domega}\\
d\beta_{kl}&:=-\Gamma_{l;k\beta}dx_\beta\label{dbeta}.
\end{eqnarray}
\end{Def}
In the literature the existence of an elastic macroscopic distortion field is generally postulated (Mura, 1987; Head et al., 1993; Cermelli and Gurtin, 2001, 2002; Koslowski et al., 2002; Ariza and Ortiz, 2005) and the global distortion decomposition in elastic and plastic parts follows\footnote{In fact, the distortion is often considered as a constitutive variable in dislocation models (Davini, 1986; Gurtin, 2002; Ariza and Ortiz, 2005).}. The point of view of the present paper is to avoid this kind of a-priori decomposition, which we believe cannot be rigorously justified. Nevertheless, the following theorem introduces rotation and distortion fields (which are not the global rotation and distortion related to the macroscopic strain) in the absence of disclinations. In contrast with the classical literature where it is basically postulated that dislocation density is the distortion curl, this relationship is here well-proved.
\begin{theorem}\label{elasticrotdist}[Bravais rotation and distortion fields]
If the macroscopic disclination density vanishes, there exists rotation and distortion fields defined as
\begin{align}
&\hspace{-20pt}\mbox{\scriptsize{BRAVAIS ROTATION}}\ &\omega_j(x)&:=\omega_j^0+\int_{x_0}^x d\omega_j,\label{omega}\\
&\hspace{-20pt}\mbox{\scriptsize{BRAVAIS DISTORTION}}\ &\beta_{kl}(x)&:=\mathcal{E}_{kl}(x^0)-\epsilon_{klj}\omega^0_j+\int_{x_0}^x d\beta_{kl},\label{beta}
\end{align}
with $\beta_{kl}=\mathcal{E}_{kl}-\epsilon_{klj}\omega_j$, and where $\omega_j^0$ is arbitrary and the integration is made on any line with endpoints $x_0$ and $x$. Moreover,
\begin{eqnarray}
\partial_\alpha\beta_{k\beta}=\eth_\alpha\eth_\beta u_k\quad\mbox{and}\quad\epsilon_{\alpha\beta}\partial_\alpha\beta_{k\beta}=\alpha_k.\label{decompmacro}
\end{eqnarray}
\end{theorem}
{\bf Proof.} 
By Definition \ref{defgeom}, the symmetric part of the connexion writes as
\begin{eqnarray}
-\Gamma_{(l;k)\beta}dx_\beta=-\frac{1}{2}\partial_\beta g_{kl}dx_\beta=-\frac{1}{2}\partial_m g_{kl}dx_m=\partial_m \mathcal{E}_{kl}dx_m=d\mathcal{E}_{kl},\nonumber
\end{eqnarray}
while, by Definition \ref{defgeom} and Proposition \ref{deltagamma}, the skew-symmetric part writes as
\begin{eqnarray}
-\Gamma_{[l;k]\beta}&=&-\frac{1}{2}(\partial_k g_{l\beta}-\partial_l g_{k\beta})+\Delta\Gamma_{l;k\beta}=\partial_k\mathcal{E}_{l\beta}-\partial_l\mathcal{E}_{k\beta}+\Delta\Gamma_{l;k\beta}.\nonumber
\end{eqnarray}
Observing, by Definitions \ref{defmacro} and \ref{formdiff} and Proposition \ref{deltagamma}, that
$\displaystyle d\omega_j=\eth_\beta\omega_j dx_\beta$ $=-\frac{1}{2}\epsilon_{lkj}\Gamma_{[l;k]\beta}dx_\beta$,
it results that $d\beta_{kl}=d\mathcal{E}_{kl}-\epsilon_{klj}d\omega_j$.
Under the assumption of a vanishing macroscopic disclination density, the existence of well-defined Bravais rotation and distortion fields follows from Eqs. (\ref{domega}) and (\ref{decompmacro}), Theorem \ref{macrodens}, and Remark \ref{stokesmacro}. Moreover, since $\partial_\alpha\beta_{k\beta}=\partial_\alpha\mathcal{E}_{k\beta}-\epsilon_{k\beta j}\eth_\alpha\omega_j$, by Eq. (\ref{transportation}), it equals $\eth_\alpha\eth_\beta u_k$, completing the proof by Eq. (\ref{dislocdensmacro}).{\hfill $\square$}
\begin{Rem}
Referring to "Bravais" instead of "elastic" rotation and distortion fields is devoted to highlight that these quantities do not have a purely elastic meaning
\end{Rem}
\begin{Rem}
The Bravais distortion does not derive from a Bravais displacement in the presence of dislocations. In fact, around a closed loop $C$, even if the macroscopic disclination density vanishes, the displacement differential as defined by $du_k:=\beta_{k\alpha}dx_\alpha$ verifies by Theorem \ref{elasticrotdist} the relationship:
\begin{eqnarray}
\int_C du_k=\int_S\epsilon_{\beta\alpha}\partial_\beta\beta_{k\alpha}dS=\alpha_k(S).
\end{eqnarray}
\end{Rem}
\begin{Rem}
Eq. (\ref{omega}) indicates that symbol $\eth$ in Eq. (\ref{domega}) becomes a true derivation operator in the absence of disclinations. 
\end{Rem}
\begin{Rem}
Theorem \ref{gamma} defines an operation of parallel displacement according to the Bravais lattice geometry. The parallel displacement of any vector $v_i$ along a curve of tangent vector $dx^{(1)}_\alpha$ is such that $dx^{(1)}_\alpha\nabla_\alpha v_i=0$ and hence that the components of $v_i$ vary according to the law $d^{(1)}v_i=-\Gamma_{i;j\beta}v_j dx^{(1)}_\beta$ (Dubrovin et al., 1992). This shows the macroscopic Burgers vector and dislocation density together with the Bravais rotation and distortion fields as reminiscences of the defective crystal properties at the nanoscale. In fact, if $dx^{(1)}_\nu, dx^{(2)}_\xi$ are two infinitesimal vectors with the associated area $dS:=\epsilon_{\nu\xi}dx^{(1)}_\nu dx^{(2)}_\xi$, it results from Eq. (\ref{torsion}), Remark \ref{stokesmacro}, and the skew symmetry of $T_{k;\alpha\beta}$ that, in the absence of disclinations, 
\begin{eqnarray}
dB_k=\alpha_k dS
=-\epsilon_{\alpha\beta}\Gamma_{k;\alpha\beta}dS
=-\Gamma_{k;\alpha\beta}(dx^{(1)}_\alpha dx^{(2)}_\beta-dx^{(1)}_\beta dx^{(2)}_\alpha)\nonumber,
\end{eqnarray}
whose right-hand side appears as a commutator verifying the relation
\begin{eqnarray}
dB_k=\epsilon_{\alpha\beta}\eth_\alpha\eth_\beta u_k dS=-\epsilon_{\alpha\beta}d^{(\alpha)}(dx^{(\beta)})\nonumber.
\end{eqnarray}
\end{Rem}
\section{Concluding remarks}\label{concl}
In this paper we have developed a $2D$ theory to analyse dislocated single crystals at the meso-scale by combining distributions with multiple-valued kinematic fields. The distributions are basically concentrated along the defect lines, which in turn form the branching lines of the multivalued fields. As a consequence of this analysis, a basic theorem relating the incompatibility tensor (as derived from the deformation field) to the Frank and Burgers vectors of the defect line has been established. This theory provides a framework for the homogenisation of the medium properties from meso- to macro-scale. In particular the macroscopic dislocation density is defined without stipulating an a-priori distorsion decomposition into elastic an plastic parts (which does not exist, actually). The classical relationship between Bravais distortion and dislocation densities, instead of being a definition, now appears as a result taking its origin from the meso-scale analysis. Moreover, the torsion and contortion tensors, which both describe the defective macroscopic crystal, are now properly understood as averages of concentrated mesoscopic tensors. Since the latter are the differentials (in an appropriate sense) of multivalued mesoscopic fields, we have shown how mesoscopic multivaluedness is recovered in the geometric properties of the non-Riemannian macroscopic crystal. In particular, in contrast with the mesoscale (where defects are due to the multivaluedness of the rotation and displacement fields) the macroscopic Burgers vector now appears as the commutator of a  non-closed differential operator related to the body torsion.
\newline Extension to the $3D$ case is under investigation. Here, the handling of non-rectilinear curves will be required in the framework of the geometric-measure theory. This should eventually make it possible to consider a set of defect curves, freely occupying the crystal with possible intersections and accumulation regions-forming so-called dislocation clusters.
\appendix
\section{Appendix: Computation of $2D$ rectilinear dislocations}\label{2Dsection}
\subsection{First group of solutions: planar displacement field}
>From the constitutive law $\sigma_{ij}^\star=\lambda\mathcal{E}_{kk}^\star\delta_{ij}+2\mu\mathcal{E}_{ij}^\star$ with $\lambda,\mu$ the Lam\'e coefficients and since $\mathcal{E}_{kk}^\star=\mathcal{E}_{\gamma\gamma}^\star$, the following planar law holds:
\begin{eqnarray}
\sigma_{\alpha\beta}^\star=\kappa^{*}\mathcal{E}_{\gamma\gamma}^\star\delta_{\alpha\beta}+2\mu\mathcal{E}^{\star D}_{\alpha\beta},\label{constitution}
\end{eqnarray}
with the planar compressibility modulus $\kappa^*$ defined by $\kappa^*:=\lambda+\mu$ and the planar deviatoric strain given by $\mathcal{E}^{\star D}_{\alpha\beta}=\mathcal{E}_{\alpha\beta}^\star-\frac{1}{2}\mathcal{E}_{\gamma\gamma}^\star\delta_{\alpha\beta}$.
>From the equilibrium conditions $\partial_{\beta}\sigma_{\beta \gamma}^\star=0$ it follows that
\begin{eqnarray}
\sigma_{\alpha\beta}^\star=\epsilon_{\alpha\gamma}\epsilon_{\beta\delta}\partial_\gamma\partial_\delta F,\label{sig}
\end{eqnarray}
for a smooth enough Airy function $F$, in such way that
\begin{eqnarray}
\sigma_{\alpha\alpha}^\star=\partial_\alpha^2 F=\Delta F.\label{tracesigma}
\end{eqnarray}
The relations between stress and strain are
\begin{eqnarray}
\mathcal{E}_{\alpha\beta}^\star=\frac{1+\nu^{*}}{E^{*}}\sigma_{\alpha\beta}^\star-\frac{\nu^{*}}{E^{*}}\sigma_{\gamma\gamma}^\star\delta_{\alpha\beta},\label{strain1}
\end{eqnarray}
with the $3D$ and planar elastic coefficients $\displaystyle E=\frac{\mu(3\lambda+2\mu)}{\lambda+\mu}$, $\displaystyle \nu=\frac{\lambda}{2(\lambda+\mu)}$, $E^{*}:=\displaystyle \frac{E}{1-\nu^{2}}$, and $\nu^*:=\displaystyle \frac{\nu}{1-\nu}$. \newline The first compatibility condition Eq. (\ref{complane}) writes from  Eqs. (\ref{tracesigma}) and (\ref{strain1}) as
\begin{eqnarray}
\Delta\Delta F=0.\nonumber
\end{eqnarray}
In this and the following sections, functions of the complex variable $Z=x+iy$ and its conjugate $\overline Z$ are now introduced.
Remembering that, compared to holomorphic functions, analytical functions may be multivalued it is easily seen that given two analytic functions $f$ and $g$, all real functions of the form 
\begin{eqnarray}
F=\Re\{\overline Z f +g\}\nonumber
\end{eqnarray}
satisfy Eq. (\ref{sig}) and vice-versa. 
Eq. (\ref{tracesigma}) then shows that  
\begin{eqnarray}
\sigma_{xx}^\star+\sigma_{yy}^\star=4\Re\{f'(Z)\}.\nonumber
\end{eqnarray} 
>From Eqs. (\ref{tracesigma}) and (\ref{strain1}) the deformation tensor is given by
\begin{eqnarray}
\left\{ \displaystyle \begin{array}{ll}\mathcal{E}_{xx}^\star+\mathcal{E}_{yy}^\star=\frac{4(1-\nu^{*})}{E^{*}}\Re\{f'(Z)\}, \\ \mathcal{E}_{yy}^\star-\mathcal{E}_{xx}^\star+2i\mathcal{E}_{xy}^\star=\frac{2(1+\nu^{*})}{E^{*}}(\overline Z f''(Z)+g''(Z)),\end{array} \right.\label{strain}
\end{eqnarray}
yielding after integration
\begin{eqnarray}
E^*(u_x^\star-iu_y^\star)&=&(3-\nu^{*})\overline f(Z)-(1+\nu^{*})(\overline Z f'(Z)+g'(Z)),\label{displ-rot_1}\\
E^*\omega_z^\star&=&4\Im\{f'(Z)\label{displ-rot_2}\}.\nonumber
\end{eqnarray}
It should be emphasised that $\mathcal{E}_{\alpha\beta}^\star$ must be a single-valued field.
\subsection{Second group of solutions: vertical displacement field}\label{second}
Another solution concerns the particular case where $u^\star_\alpha=0$, noting that each solution of $2D$ elasticity can be decomposed into a purely planar and a purely vertical solution. In fact, since stress equilibrium shows that \[(\lambda+\mu)\partial_i\partial_j u_j^\star+\mu\Delta u_i^\star = 0,\] it is easy to infer for $i=z$ that 
\begin{eqnarray}
u_z^\star=\frac{(1+\nu)}{E}\Re\{{h(Z)}\},\label{displvert}
\end{eqnarray}
with $h(Z)$ an analytic function. Then
\begin{eqnarray}
\mathcal{E}_{xz}^\star-i\mathcal{E}_{yz}^\star=\frac{(1+\nu)}{2E}h'(Z).\label{strainvert}
\end{eqnarray}
The function $h'(Z)$ must be uniform. The complex rotation is
\begin{eqnarray}
\omega^\star:=\omega_x^\star+i\omega_y^\star=-\frac{i(1+\nu)}{2E}\overline{h'}(Z).\label{rotvert}
\end{eqnarray}
In $2D$ isothermal linear elasticity without body forces, every displacement solution has planar components given by Eq. (\ref{displ-rot_1}) and a vertical component given by Eq. (\ref{displvert}) while the rotation vector has planar components given by Eq. (\ref{rotvert}) and a vertical component given by Eq. (\ref{displ-rot_2}) (Sokolnikoff (1946) and Knopp (1996)).
\subsection{The three $2D$ examples of rectilinear defects}\label{examples}
In this section we consider two typical multivalued analytic functions $log(Z)$ and $Zlog(Z)$. Starting from the general uniform strain expressions Eq. (\ref{strain}) or Eq. (\ref{strainvert}) it is easily observed that any of the holomorphic functions $f''$ (with $\Re{\{f'\}}$ single-valued), $g''$ and $h'$ can provide a solution to the $2D$ problem. Since these functions can be expanded in Laurent series:
\begin{eqnarray}
f''(Z)=\displaystyle\sum_{-\infty}^{+\infty}a_nZ^n \quad,\quad g''(Z)=\displaystyle\sum_{-\infty}^{+\infty}b_nZ^n \quad,\quad h'(Z)=\displaystyle\sum_{-\infty}^{+\infty}c_nZ^n,\nonumber
\end{eqnarray}
inside their respective convergence annuli, primitivation shows that
\begin{displaymath}
\left \{\begin{array}{lll}  f(Z)=\displaystyle\sum_{\substack{-\infty\\
 n\not=-1,-2}}^{+\infty}\frac{a_n}{{(n+1)}{(n+2)}}Z^{n+2}-a_{-2}\ln(Z)+a_{-1}Z\ln(Z)+A_1
Z+A_0,\\ g(Z)=\displaystyle\sum_{\substack{-\infty\\ n\not= -1,-2}}^{+\infty}\frac{b_n}{{(n+1)}{(n+2)}}Z^{n+2}-b_{-2}\ln(Z)+b_{-1}Z\ln(Z)+B_1Z +B_0, \\h(Z)=\displaystyle\sum_{\substack{-\infty\\ n\not=-1}}^{+\infty}\frac{c_n}{n+1}Z^{n+1}+c_{-1}\ln(Z)+C_0,
 \end{array} \right.
\end{displaymath}
with $a_{-1}$ real in order that $\Re{\{f'\}}$ be uniform. The relevant cases are those which give rise to a dislocation or a disclination, i.e. such that the functions $\overline f,f',g',\Re{\{h\}}$ or $\Im{\{f'\}}$ are multivalued. Hence, in order to obtain non-vanishing rotation or displacement jumps, one needs to consider the following cases:
\begin{eqnarray}
f(Z)&=&-a_{-2}\ln(\frac{Z}{R})+a_{-1}Z\ln(\frac{Z}{R}), \quad a_{-1}\in\R,\label{relevant1}\\
g(Z)&=&b_{-1}Z\ln(\frac{Z}{R}),\label{relevant2} \\
h(Z)&=&c_{-1}\ln(\frac{Z}{R}),\qquad c_{-1}\in i\R,\label{relevant3}
\end{eqnarray}
where $R$ is a constant length and to which any purely elastic term may always be added. In fact, from Eqs. (\ref{displ-rot_1}), (\ref{displ-rot_2}), (\ref{displvert}) and (\ref{rotvert}), and from the definition $B^\star:=B_x^\star+iB_y^\star$ with $B_k^\star$ given by Eq. (\ref{burgers}),
it follows that:
\begin{eqnarray}
\left \{\begin{array}{lll}  \Om_z^\star&=&\displaystyle \frac{4}{E^*}[\Im\{f'\}],
\\[8pt]\overline{B}^\star:&=&B_x^\star-iB_y^\star=[u_x^\star]-i[u_y]-\Om_z^\star (iz)
\\\displaystyle&=&\frac{3-\nu^*}{E^*}[\overline{f}]-\frac{1+\nu^*}{E^*}\left\{\overline{Z}[f']+[g']\right\}+\frac{4i\overline{Z}}{E^*}[\Im\{f'\}],
\\[8pt]\Om^\star:&=&\Om_x^\star+i\Om_y^\star=\displaystyle -\frac{i(1+\nu)}{2E}[\overline{h'}],
\\[8pt]B_z^\star&=&[u_z^\star]-\Re\{i\overline{Z}\Om^\star\}=\displaystyle \frac{1+\nu}{E}[\Re\{h\}]-\frac{1+\nu}{2E}\Re\{\overline{Z}[\overline{h'}]\}=\displaystyle \frac{1+\nu}{E}[\Re\{h\}].
 \end{array} \right. \nonumber
\end{eqnarray}
It should immediately be noted that $\Om^\star$ vanishes identically since $h'$ cannot be multivalued.
>From Eqs. (\ref{relevant1})-(\ref{relevant3}) and some easy computations, the only possible solutions are given by the following proposition.
\begin{proposition}\label{solutions}
For a straight defect line $L$ in $2D$ elasticity, there are no more than three distinct defect classes. The two dislocation classes are the screw dislocation with a vertical Burgers vectors $B^\star_z$, as generated by the analytical function $h$ ($f=g=0$), and the edge dislocation with a planar complex Burgers vector $B^\star_x+iB^\star_y$, as generated by the analytical function $g$ ($f=h=0$). There is a single class of disclinations, the wedge disclination, which has a vertical Frank vector $\Om^\star_z$ and is generated by the analytical function $f$ ($g=h=0$). These functions are:
\begin{eqnarray}
&&\hspace{-65pt}\mbox{\scriptsize WEDGE DISCLINATION}\hspace{40pt}f(Z)=\frac{E^*\Om^\star_z}{8\pi}z\ln(\frac{Z}{R})\nonumber\\
&&\hspace{-65pt}\mbox{\scriptsize EDGE DISLOCATION}\hspace{52pt}g(Z)=\frac{E^*(B^\star_y+iB^\star_x)}{2(1+\nu^*)\pi}Z\ln(\frac{Z}{R})\nonumber\\
&&\hspace{-65pt}\mbox{\scriptsize SCREW DISLOCATION}\hspace{45pt}h(Z)=\frac{iEB^\star_z}{2\pi(1+\nu)}\ln(\frac{Z}{R}).\nonumber
\end{eqnarray}
\end{proposition}
For the edge dislocation, a detailed derivation is given by Eshelby (1966).

\end{document}